\documentclass[twoside,twocolumn,9pt]{article}
\usepackage{extsizes}
\usepackage[super,sort&compress,comma]{natbib} 
\usepackage[version=3]{mhchem}
\usepackage[left=1.5cm, right=1.5cm, top=1.785cm, bottom=2.0cm]{geometry}
\usepackage{balance}
\usepackage{times,mathptmx}
\usepackage{sectsty}
\usepackage{graphicx} 
\usepackage{lastpage}
\usepackage[format=plain,justification=justified,singlelinecheck=false,font={stretch=1.125,small,sf},labelfont=bf,labelsep=space]{caption}
\usepackage{float}
\usepackage{fancyhdr}
\usepackage{fnpos}
\usepackage[english]{babel}
\addto{\captionsenglish}{%
  
}
\usepackage{array}
\usepackage{droidsans}
\usepackage{charter}
\usepackage[T1]{fontenc}
\usepackage[usenames,dvipsnames]{xcolor}
\usepackage{setspace}
\usepackage[compact]{titlesec}
\usepackage{hyperref}

\usepackage{epstopdf}

\usepackage{lipsum}
\usepackage{multicol}

\usepackage{cuted}
\usepackage{amsmath}

\definecolor{cream}{RGB}{222,217,201}

\begin{document}

\pagestyle{fancy}
\thispagestyle{plain}
\fancypagestyle{plain}{

\fancyhead[C]{\includegraphics[width=18.5cm]{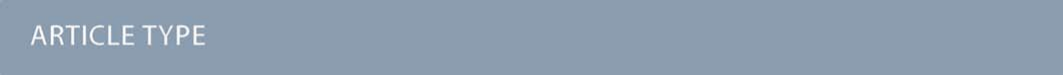}}
\fancyhead[L]{\hspace{0cm}\vspace{1.5cm}\includegraphics[height=30pt]{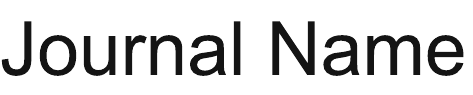}}
\fancyhead[R]{\hspace{0cm}\vspace{1.7cm}\includegraphics[height=55pt]{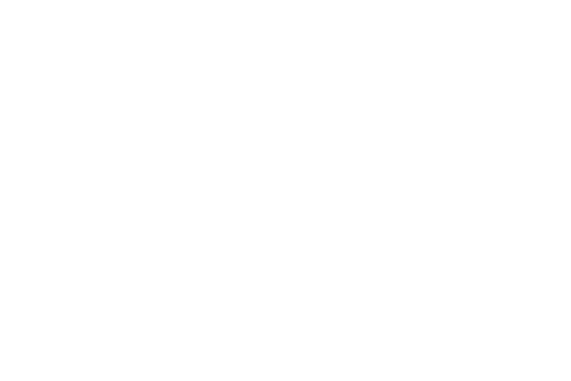}}
\renewcommand{\headrulewidth}{0pt}
}

\makeFNbottom
\makeatletter
\renewcommand\LARGE{\@setfontsize\LARGE{15pt}{17}}
\renewcommand\Large{\@setfontsize\Large{12pt}{14}}
\renewcommand\large{\@setfontsize\large{10pt}{12}}
\renewcommand\footnotesize{\@setfontsize\footnotesize{7pt}{10}}
\makeatother

\renewcommand{\thefootnote}{\fnsymbol{footnote}}
\renewcommand\footnoterule{\vspace*{1pt}%
\color{cream}\hrule width 3.5in height 0.4pt \color{black}\vspace*{5pt}} 
\setcounter{secnumdepth}{5}

\makeatletter 
\renewcommand\@biblabel[1]{#1}            
\renewcommand\@makefntext[1]%
{\noindent\makebox[0pt][r]{\@thefnmark\,}#1}
\makeatother 
\renewcommand{\figurename}{\small{Fig.}~}
\sectionfont{\sffamily\Large}
\subsectionfont{\normalsize}
\subsubsectionfont{\bf}
\setstretch{1.125} 
\setlength{\skip\footins}{0.8cm}
\setlength{\footnotesep}{0.25cm}
\setlength{\jot}{10pt}
\titlespacing*{\section}{0pt}{4pt}{4pt}
\titlespacing*{\subsection}{0pt}{15pt}{1pt}

\fancyfoot{}
\fancyfoot[LO,RE]{\vspace{-7.1pt}\includegraphics[height=9pt]{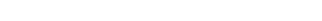}}
\fancyfoot[CO]{\vspace{-7.1pt}\hspace{13.2cm}\includegraphics{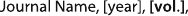}}
\fancyfoot[CE]{\vspace{-7.2pt}\hspace{-14.2cm}\includegraphics{head_foot/RF}}
\fancyfoot[RO]{\footnotesize{\sffamily{1--\pageref{LastPage} ~\textbar  \hspace{2pt}\thepage}}}
\fancyfoot[LE]{\footnotesize{\sffamily{\thepage~\textbar\hspace{3.45cm} 1--\pageref{LastPage}}}}
\fancyhead{}
\renewcommand{\headrulewidth}{0pt} 
\renewcommand{\footrulewidth}{0pt}
\setlength{\arrayrulewidth}{1pt}
\setlength{\columnsep}{6.5mm}
\setlength\bibsep{1pt}


\newcommand{\beq}{\begin{equation}}
\newcommand{\eeq}{\end{equation}}

\newcommand{\be}{\begin{equation}} 
\newcommand{\ee}{\end{equation}}
\newcommand{\bea}{\begin{eqnarray}}   
\newcommand{\eea}{\end{eqnarray}}

\newcommand{\br}{{\bf r}}
\newcommand{\bu}{{\bf u}}
\newcommand{\bv}{{\bf v}}
\newcommand{\bR}{{\bf R}}
\newcommand{\bRz}{{\bf R}^0}
\newcommand{\bk}{{ \bf k}}
\newcommand{\bx}{{ \bf x}}
\newcommand{\vv}{{\bf v}}
\newcommand{\bn}{{\bf n}}
\newcommand{\mb}{{\bf m}}
\newcommand{\bq}{{\bf q}}
\newcommand{\bG}{{\bf G}}
\newcommand{\rb}{{\bar r}}
\newcommand{\rr}{{\bf r}}

\newcommand{\kk}{\boldsymbol{\kappa}}
\newcommand{\greeketabold}{\boldsymbol{\eta}}
\newcommand{\xxi}{\boldsymbol{\xi}}
\newcommand{\cchi}{\boldsymbol{\chi}}
\newcommand{\bomega}{\boldsymbol{\Omega}}
\newcommand{\Tr}{\boldsymbol{Tr}}


\makeatletter 
\newlength{\figrulesep} 
\setlength{\figrulesep}{0.5\textfloatsep} 

\newcommand{\topfigrule}{\vspace*{-1pt}%
\noindent{\color{cream}\rule[-\figrulesep]{\columnwidth}{1.5pt}} }

\newcommand{\botfigrule}{\vspace*{-2pt}%
\noindent{\color{cream}\rule[\figrulesep]{\columnwidth}{1.5pt}} }

\newcommand{\dblfigrule}{\vspace*{-1pt}%
\noindent{\color{cream}\rule[-\figrulesep]{\textwidth}{1.5pt}} }

\makeatother

\twocolumn[
  \begin{@twocolumnfalse}
\vspace{3cm}
\sffamily
\begin{tabular}{m{4.5cm} p{13.5cm} }

\includegraphics{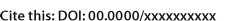} & \noindent\LARGE{\textbf{Spontaneous generation of angular momentum in chiral active crystals$^\dag$}} \\
\vspace{0.3cm} & \vspace{0.3cm} \\

 & \noindent\large{Umberto Marini Bettolo Marconi \textit{$^{a}$} and Lorenzo Caprini $^{\ast}$\textit{$^{b}$} } \\

\includegraphics{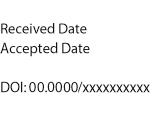} & \noindent\normalsize{

We study a two-dimensional chiral active crystal composed of underdamped chiral active particles. These particles, characterized by intrinsic handedness and persistence, interact via linear forces derived from harmonic potentials. Chirality plays a pivotal role in shaping the system's behavior: it reduces displacement and velocity fluctuations while inducing cross-spatial correlations among different Cartesian components of velocity. These features distinguish chiral crystals from their non-chiral counterparts, leading to the emergence of net angular momentum, as predicted analytically. This angular momentum, driven by the torque generated by the chiral active force, exhibits a non-monotonic dependence on the degree of chirality. Additionally, it contributes to the entropy production rate, as revealed through a path-integral analysis.
We investigate the dynamic properties of the crystal in both Fourier and real space. Chirality induces a non-dispersive peak in the displacement spectrum, which underlies the generation of angular momentum and oscillations in time-dependent autocorrelation functions or mean-square displacement, all of which are analytically predicted.
}\\

%
%

\end{tabular}

 \end{@twocolumnfalse} \vspace{0.6cm}

  ]

\renewcommand*\rmdefault{bch}\normalfont\upshape
\rmfamily
\section*{}
\vspace{-1cm}


\footnotetext{\textit{$^{a}$~School of Sciences and Technology, University of Camerino, Via Madonna delle Carceri, I-62032, Camerino, Italy. }}
\footnotetext{\textit{$^{b}$~Sapienza University of Rome, Piazzale Aldo Moro 2, Rome, Italy. E-mail: lorenzo.caprini@gssi.it, and lorenzo.caprini@uniroma1.it}}


\footnotetext{The authors contributed equally to this work.}


\section{Introduction}

\begin{figure*}[!t]
\centering
\includegraphics[width=1\linewidth,keepaspectratio]{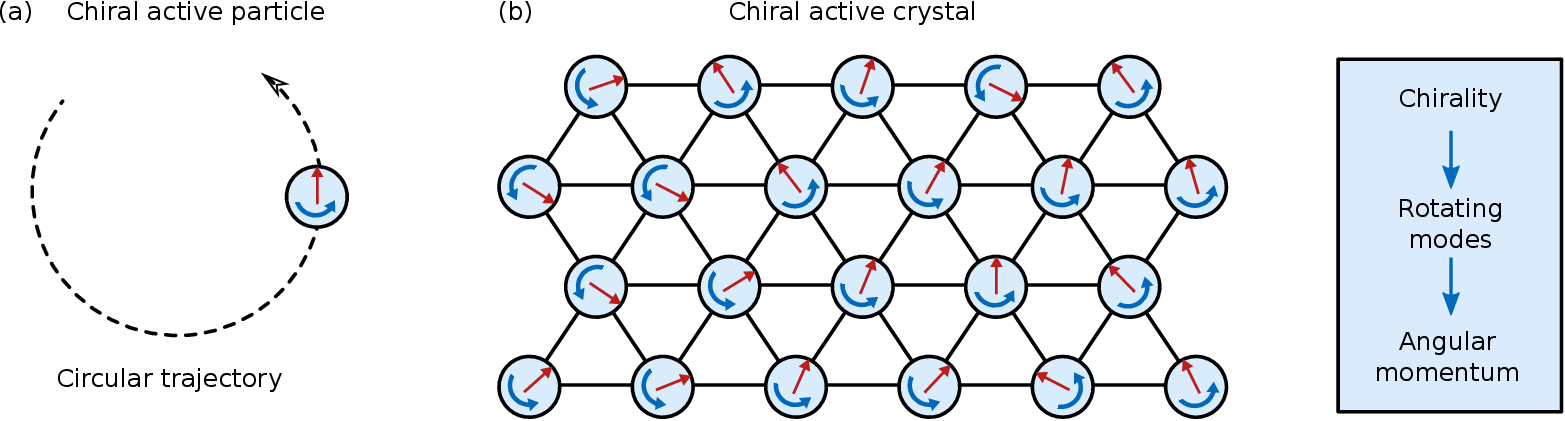}
\caption{
{\textbf{Chiral active crystals.}}
(a) Illustration of a chiral active particles which self-propel (red arrow) and rotate counterclockwise (blue arrow) performing a circular trajectory.
(b) Illustration of a chiral active crystal in two dimensions in a triangular lattice.
Chirality induces rotating modes (a non-dispersive peak in the displacement spectrum) which generates a global angular momentum.
}\label{fig:Fig1_presentation}
\end{figure*}

The term ``chirality'', derived from the ancient Greek word ``$\chi \epsilon \iota \rho$'' meaning ``hand'', was coined by Lord Kelvin in 1873 to describe the property of an object that cannot be superimposed onto its mirror image~\cite{pal2020chirality}. Chirality, also known as handedness, refers to the characteristic of being either right-handed or left-handed. This property is not exclusive to humans but is observed across various scales in nature, from organic molecules like DNA and proteins to different organisms such as plants and animals. For instance, snails exhibit chirality, with some possessing right-handed spiraling shells while others have left-handed shells or spiral in a clockwise/counterclockwise direction~\cite{sinjushin2019molecular}. Chirality is also evident in the spiral arrangement of leaves, stems, roots, and floral parts in plants.

Chirality has found applications in active matter~\cite{marchetti2013hydrodynamics, elgeti2015physics, bechinger2016active}, where materials are engineered to exhibit specific behaviors, offering opportunities for designing structures with diverse functionalities. As a result, there has been increasing interest in studying the behavior of chiral self-propelled particles~\cite{lowen2016chirality, liebchen2022chiral}, both synthetic and natural, to achieve control over their motion. These particles exhibit two distinctive properties: first, they break time-reversal symmetry, converting energy into directed motion and dissipated heat; second, they break reflection symmetry, inherently existing far from equilibrium.
The breaking of time-reversal symmetry gives rise to phenomena absent in equilibrium systems, such as flocking~\cite{vicsek2012collective, cavagna2014bird}, motility-induced phase separation~\cite{cates2015motility, mandal2019motility, van2019interrupted, caporusso2020motility, caprini2022role}, spatial velocity correlations~\cite{caprini2020spontaneous, henkes2020dense, caprini2021spatial, szamel2021long, keta2024emerging, abbaspour2024long}, and accumulation near obstacles~\cite{uspal2015self, wittmann2016active, caprini2018active, das2018confined}. In contrast, the breaking of reflection symmetry results in distinct phenomena, including odd diffusivity~\cite{hargus2021odd, kalz2022collisions, kalz2024oscillatory} and odd viscosity~\cite{banerjee2017odd, han2021fluctuating, lou2022odd, fruchart2022odd}. A prominent example of broken parity is the Hall effect, where moving charges in a conductor subjected to a magnetic field produce an electric potential difference perpendicular to both the motion and the field, manifesting as an off-diagonal Hall resistance.
Focusing on active matter, numerous examples of chirality are found in both natural systems and synthetic structures. Observing circular swimming in two dimensions or helical swimming in three dimensions requires only a slight asymmetry in the active particles relative to their propulsion axis~\cite{kummel2013circular, monderkamp2023network} or the presence of an external magnetic field~\cite{cao2023memory}. For instance, Escherichia coli bacteria and spermatozoa in suspensions exhibit spiral-like swimming paths or circular motion near planar substrates~\cite{lauga2006swimming, petroff2015fast, perez2019bacteria}. Similarly, magnetotactic bacteria, which are motile prokaryotes equipped with magnetosome organelles, act like miniature compasses, swimming along magnetic field lines. In addition, droplets of cholesteric liquid crystals in isotropic liquids~\cite{carenza2019rotation} and artificial self-propelled L-shaped particles~\cite{kummel2013circular} also display circular trajectories.

The behavior of individual chiral active particles is well understood: the angular drift leads to circular trajectories~\cite{van2008dynamics} (see Fig.\ref{fig:Fig1_presentation}(a)), resulting in temporal oscillations in velocity autocorrelations and mean-square displacement, as well as a reduction in long-time diffusion~\cite{van2008dynamics, sevilla2016diffusion, reichhardt2019active, chepizhko2020random, han2021fluctuating, van2022role, olsen2021diffusion, shi2024chiral}.
Beyond the case of a potential-free particle, the impact of chirality has been investigated in various scenarios: dimers composed of two chiral particles~\cite{muzzeddu2022active}, harmonic confinement where chirality reduces the effective temperature induced by activity~\cite{caprini2023chiral}, and near planar obstacles where circular motion mitigates the wall accumulation typically observed in active particles~\cite{caprini2019activechiral, fazli2021active, mecke2024chiral, valecha2024chirality}.

Moreover, the impact of chirality on collective phenomena has been extensively investigated, both in the presence and absence of alignment interactions.
In systems with alignment interactions, chirality gives rise to pattern formation~\cite{liebchen2017collective, levis2018micro, ai2018mixing, kruk2020traveling, liao2021emergent, negi2022geometry, zhang2022collective, kreienkamp2022clustering, hiraiwa2022collision, lei2023collective, negi2023geometry, ceron2023diverse}, including phenomena such as chiral self-recognition~\cite{arora2021emergent}, traveling waves~\cite{liebchen2016pattern}, and rotating micro-flock patterns~\cite{liebchen2017collective}.
In the absence of alignment interactions, chirality suppresses motility-induced phase separation~\cite{liao2018clustering, ma2022dynamical, semwal2022macro, sese2022microscopic, bickmann2022analytical} and induces a hyperuniform phase~\cite{lei2019nonequilibrium, huang2021circular, zhang2022hyperuniform, Kuroda_2023, kuroda2024long}. Additionally, circular motion reduces spatial velocity correlations by decreasing their correlation length~\cite{shee2024emergent}.
In this case, chirality appears to primarily diminish the distinctive effects characteristic of active matter. However, recent studies have highlighted novel phenomena that are primarily associated with rotational dynamics arising from circular motion. These effects include unique oscillatory caging phenomena in chiral active glasses~\cite{debets2023glassy}, fascinating vortex patterns in velocity fields~\cite{liao2018clustering, zhang2020reconfigurable}, self-reverting vorticity in the presence of attractive interactions~\cite{caprini2024self}, and demixing in binary mixtures~\cite{lauga2006swimming, petroff2015fast, perez2019bacteria}.

Understanding the role of activity and chirality in active crystals remains a challenge. 
In the absence of lattice vacancies, the particles in a solid structure are restricted to vibrate around their lattice positions rather than diffuse. In one- and two-dimensional systems, these vibrational excursions can be substantial, even diverging for infinite systems. While the physics of non-chiral solids has been extensively studied, the inclusion of chirality introduces additional complexity.
Recently, this problem was addressed by Shee et al.~\cite{shee2024emergent}, who employed a continuum theory. Their study predicts that chiral active crystals exhibit spatial velocity correlations following an Ornstein-Zernike profile, as seen in non-chiral active solids~\cite{henkes2020dense, caprini2020hidden, keta2022disordered}, with a correlation length that decreases as chirality increases.

In our paper, we investigate chiral active crystals using a particle-based approach, previously applied to non-chiral crystals~\cite{caprini2023entropons} (see Fig.\ref{fig:Fig1_presentation}(b) for an illustration). For harmonic crystals, this method enables us to analytically derive the displacement spectrum without the need for parameter fitting, as well as to make approximations for predicting spatiotemporal correlations in real space.
After validating the findings of Ref.~\cite{shee2024emergent} on spatial velocity correlations, we discover that chirality induces spatial structures in the cross-velocity correlations, involving mixed Cartesian components. This phenomenon is associated with a net angular momentum entirely driven by circular motion, which displays a non-monotonic dependence on chirality.
The analytical solution for the displacement spectrum reveals the presence of a non-dispersive peak at the chirality frequency, which underpins temporal oscillations in displacement autocorrelations and mean-square displacements. Interestingly, chirality introduces an additional contribution to entropy production, arising from the torque exerted by the active forces on the particles of the crystal.

The paper is organized as follows: Section~\ref{Model} introduces the model, presenting an extension of the Active Ornstein-Uhlenbeck model to describe a two-dimensional chiral active crystal. Section~\ref{sec:dynamicalcorrelations} examines the model's dynamical correlations in the frequency and wave vector domains, including displacement correlations (both diagonal and off-diagonal) and angular momentum.
In Sec.~\ref{sec:staticcorrelations}, we analyze the steady-state properties of the system, such as equal-time displacement-displacement and velocity-velocity correlations, angular momentum, and entropy production, while in Sec.~\ref{timedependent} we explore the system's temporal behavior, including the mean-square displacement and the two-time correlations of velocity and displacement.
Finally, Sec.~\ref{sec:conclusions} presents the conclusions. The paper also includes several appendices that detail technical aspects omitted from the main text for clarity and brevity.

\section{Model for chiral active crystals}
\label{Model}

We consider a two-dimensional non-equilibrium crystal composed of chiral active particles in the underdamped regime. Each particle is harmonically connected to its neighbors and subjected to an additional chiral active force that induces clockwise or counterclockwise rotations.
Specifically, the particles are distinguishable and labeled by a two-dimensional index  $\bn=(n_1,n_2)$.
They oscillate around their equilibrium lattice positions, $\bR_\bn$,
under the influence of random, active, and conservative forces governing their dynamics.
The instantaneous coordinates  $\br_\bn(t)$  
of the particles are described by the following set of coupled Langevin equations:
\begin{eqnarray}
&&
m\ddot \br_\bn(t) = -m\gamma \dot \br_\bn(t)
+ \mathbf{F}_n + {\bf f}^a_\bn(t) +  \sqrt{2 m\gamma T} \xxi_\bn(t)  \,,
 \label{dynamicequation0}
\end{eqnarray}
where $\boldsymbol{\zeta}_\bn(t)$ is a white noise with zero mean and unit variance.

The term $\gamma$ is a friction coefficient per unit  mass, such that the inertial time is simply $1/\gamma$. The constant $T$ sets the amplitude of the noise, which corresponds to the solvent temperature in the case of active colloids or it is generated by plate and particle imperfections in active granular particles~\cite{scholz2018inertial}.
The term $\mathbf{F}_n=-\sum_\mb^{n.n}\nabla_n U(|\br_\mb-\br_\bn|)$ represents the conservative force at position $\br_\bn$ arising from a potential $U$ ensuring that the particle is maintained at the lattice spacing $\sigma$. Assuming short-range interactions, the sum $\sum^{n.n}$ is taken over the nearest neighbors of the particle which at equilibrium occupies the lattice position, $\bR_\bn$.
The term $\mathbf{f}^a_n$ represents the active or self-propelled force driving each particle out of equilibrium.
The evolution of $\mathbf{f}^a_n$ follows the active Ornstein-Uhlenbeck particle (AOUP) dynamics~\cite{berthier2019glassy, wittmann2018effective, martin2021statistical} extended to include chirality~\cite{caprini2019activechiral}, and is expressed as
\begin{eqnarray}
&&
\dot {\bf f}^a_\bn(t) =-\frac{1}{\tau} {\bf f}^a _\bn(t)+\bomega\times {\bf f}^a _\bn(t)+m\gamma v_0 \sqrt{\frac{2 } { \tau}}  \cchi_\bn(t)\,,
\label{eq:activeforce}
\end{eqnarray}
where $\boldsymbol{\chi}_\bn(t)$ is a white noise with zero mean and unit variance.

The terms $v_0$ and $\tau$ represent the typical particle speed and the persistence time, respectively, i.e.\ the time required for a particle to randomize its velocity.
Following Ref.~\cite{caprini2019activechiral}, chirality is incorporated into the dynamics by introducing the term $\bomega\times {\bf f}^a _\bn$, where $\bomega=\Omega {\bf z}$ is a vector perpendicular to the plane of motion with a magnitude of 
 $\Omega$ (${\bf z}$ is the unit vector in the vertical direction). The parameter $\Omega$ is an effective constant torque and is commonly  referred to as chirality, determining the revolution time of  ${\bf f}^a _\bn$.
 This model can be interpreted as a Gaussian approximation of chiral active Brownian motion. It reproduces the temporal correlation functions of ${\bf f}^a _\bn$ in the steady state, expressed as follows:
\begin{subequations}
\begin{align}
&\langle f_{\bn}^{a,x}(t) f_{\bn'}^{a,x}(t')\rangle
=  m^2\gamma^2 v_0^2 \, \delta_{\bn,\bn'} e^{-|t-t'|/\tau}
\cos(\Omega (t-t'))
\label{exex}
\\
&
\langle f_\bn^{a,x}(t) f_{\bn'}^{a,y}(t')\rangle 
=-  m^2\gamma^2 v_0^2 \, \delta_{\bn,\bn'}  e^{-|t-t'|/\tau} \sin(\Omega |t-t'|)
\label{exey}\\
&
\langle f_\bn^{a,y}(t) f_{\bn'}^{a,x}(t')\rangle  =-\langle f_\bn^{a,x}(t) f_{\bn'}^{a,y}(t')\rangle 
\label{e_cross}\,,
\end{align}
\end{subequations}
where $\langle f_{\bn}^{a,x}(t) f_{\bn'}^{a,x}(t')\rangle =\langle f_\bn^{a,y}(t) f_{\bn'}^{a,y}(t')\rangle$.
The active forces act independently on different particles, as shown by the presence of the Kronecker symbols.
As is typical for non-chiral active particles, the persistence time $\tau$ corresponds to the decay time of the temporal autocorrelations.
Chirality introduces two primary effects:\\
i) It induces oscillations with frequency $\Omega$ and
ii)  gives rise to cross-temporal correlations, i.e., correlations between different Cartesian components of the active force.
These cross-correlations vanish as $\Omega \to 0$ and exhibit odd symmetry under the exchange of $x$ and $y$, as evident from Eq.~\eqref{e_cross}. This property underpins the odd diffusivity, an antisymmetric diffusion tensor, which has been previously predicted for chiral active particles~\cite{hargus2021odd}. 
Assuming harmonic interactions among nearest neighbor particles and introducing the displacement $\bu_\bn$ with respect to their equilibrium position as $\bu_\bn=\br_\bn-\bR_\bn$,
we focus on the dynamics of the displacement vectors
\begin{eqnarray}
\ddot \bu_\bn(t)+
\gamma \dot \bu_\bn(t) = \frac{K}{m} \sum_\mb^{n.n} (\bu_\mb-\bu_\bn)+ \frac{{\bf f}^a _\bn(t)}{m}  + \sqrt{\frac{ 2 \gamma T}{m}} \xxi_n(t)  
 \label{dynamicequation2}
\end{eqnarray}
where $K$ is the strength of the elastic restoring force.
In the case of a non-linear potential $K$ 
can be determined by differentiating the potential
$U(r)$, such that $K=U''(\sigma)+U'(\sigma)/\sigma$, where $\sigma$ is the lattice spacing.

\section{Dynamical correlations}
\label{sec:dynamicalcorrelations}
The dynamics~\eqref{dynamicequation2} can be solved by switching to normal coordinates and applying the double Fourier-transform of the variables  moving from the discrete space of coordinates indexed by $\bf{n}$ and the time domain $t$ to the discrete wave vector $\bq=(q_x,q_y)$ and frequency $\omega$. 
 Specifically, we use the continuum Fourier transform to handle the time domain and the discrete Fourier transform for the spatial coordinates.
In this way, we obtain
\begin{equation}
\tilde{\bu}_{\bq}(\omega)=\frac{1}{\sqrt N}\sum_{\bn=1}^N  e^{i \bq \cdot \bR_\bn}\int_{-\infty}^{\infty} dt\, e^{i\omega t} \bu_\bn(t) \,.
\label{eq:FTdef}
\end{equation}
where the double Fourier transforms of any variables are indicated by a 'tilde' symbol and by the explicit dependence on $\bq$ and $\omega$.
The same definition can be applied to the velocity, active force, and noise variables. 
In addition, a 'hat' symbol denotes the spatial Fourier transform of a variable and a 'bar' symbol the Fourier transform in frequency
(see Appendix \ref{definitionsfouriertransform}).
Since
the velocity variable is given $\hat{\mathbf{v}}_{\bq}(t) =\hat{\dot{\mathbf{u}}}_{\bq}(t)$,  its double Fourier transform satisfies the relation $\tilde{\mathbf{v}}_{\bq}(\omega) =i\omega \tilde{\mathbf{u}}_{\bq}(\omega)$.
By applying the time and space Fourier transform to Eq.~\eqref{dynamicequation2}, we obtain the evolution of each Fourier mode of the displacement
\begin{equation}
\tilde \bu_{\bq}(\omega)=\tilde {\boldsymbol{\mathcal{R}}}^{\hat{u}\hat{u}}_\mathbf{q} ( \omega) \cdot\Bigl( \frac{\tilde {\bf f}^a_{\bq}(\omega)}{m}+\sqrt{ \frac{2\gamma T}{m  }}\tilde \xxi_{\bq}(\omega)\Bigr) \,,
\label{uqomegaequation}
\end{equation}
where we have introduced the matrix $\boldsymbol{\mathcal{R}}^{\tilde{u}\tilde{u}}_\mathbf{q}( \omega))=G_{\bq}(\omega) \mathbf{I}$ written as the product between the identiy matrix $I$ and the propagator $G_{\bq}(\omega)$ given by
\begin{equation}
\tilde G_{\bq}(\omega) =\frac{1}{\Bigl(-\omega^2+i \omega\gamma  +  \omega_{\bq}^2 \Bigr) }\,.
\label{underdampG}
\end{equation}
Here, $\omega_q$ is the frequency of the vibrational modes of the crystal, which for a triangular lattice with lattice constant $\sigma$, reads:
\begin{flalign}
\omega_{\bq}^2&=2 \frac{K}{m} \left[3 -\cos(q_x \sigma) -2\cos\left(\frac{1} {2} q_x \sigma\right)\cos \left(\frac{\sqrt 3} {2} q_y \sigma\right)\right] \nonumber\\
&\approx \frac{K}{m}\frac{3}{2}\mathbf{q}^2\sigma^2 = c^2 \mathbf{q}^2\,.
\end{flalign} 
In the last approximation, we have expanded $\omega_{\bq}$ for small wave vectors $\mathbf{q}$, obtaining the usual $\bq^2$ behavior and we have introduced $c$, as  the equivalent of the speed of sound.

\subsection{Dynamical correlations of the particle displacement}

\begin{figure*}[!t]
\centering
\includegraphics[width=1\linewidth,keepaspectratio]{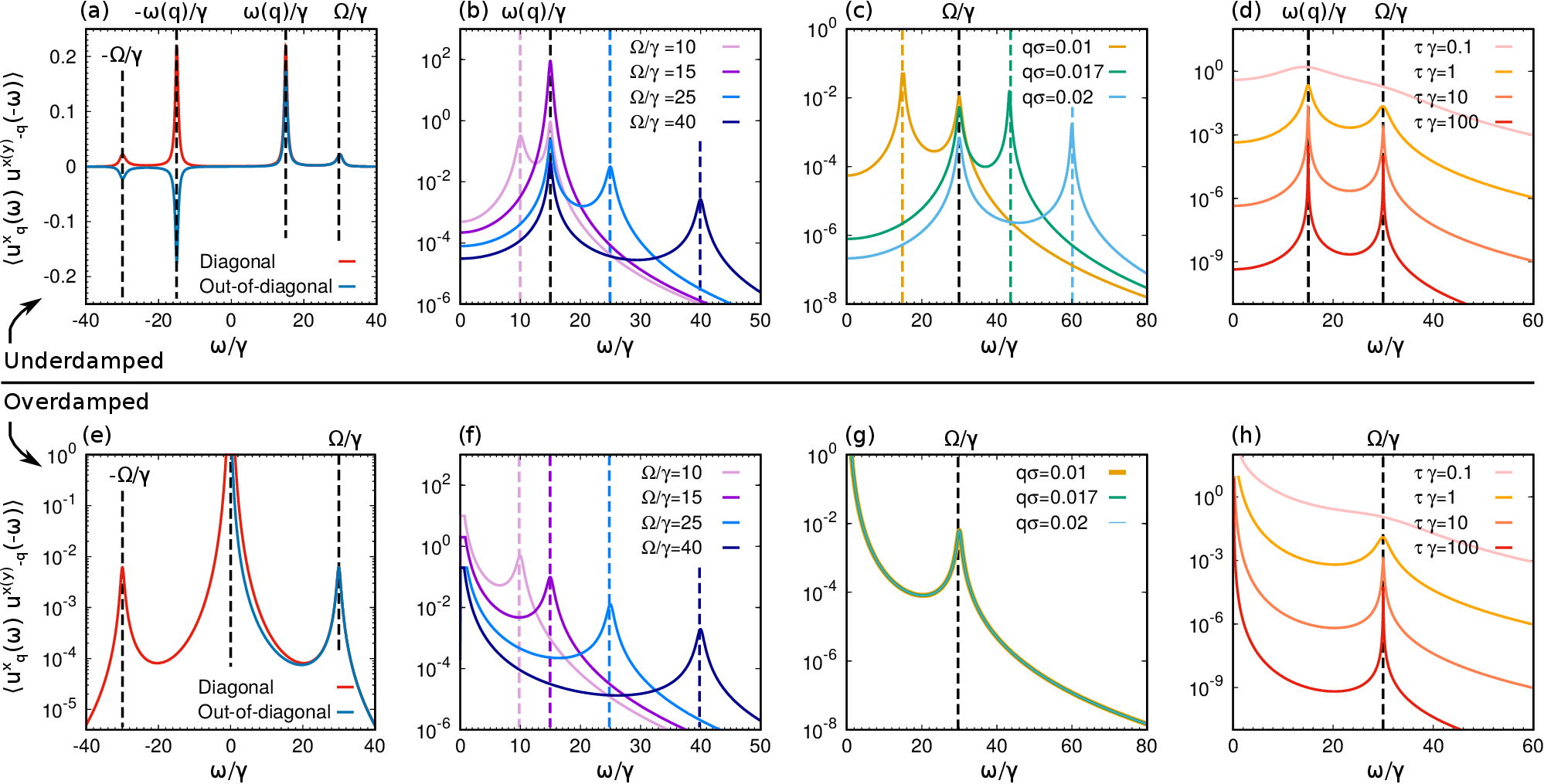}
\caption{
{\textbf{Dynamical correlation of the displacement.}}
(a), (e) Diagonal correlations of the particle displacement (normalized with the lattice constant $\sigma$), $\langle u^x_\bq(\omega) u^x_{-\bq}(-\omega) \rangle$ (red) and imaginary part of the out-of-diagonal correlations of the particle displacement $\langle u^x_\bq(\omega) u^y_{-\bq}(-\omega) \rangle$ (blue). These observables are plotted as a function of the frequency $\omega$ rescaled with the inertial time $1/\gamma$ at wave number modulus $q=|\bq|$. 
(b), (f) $\langle u^x_{\bq}(\omega) u^x_{-\bq}(-\omega) \rangle$ for different chirality values $\Omega/\gamma$.
(c), (g) $\langle u^x_{\bq}(\omega) u^x_{-\bq}(-\omega) \rangle$ for different rescaled wave numbers $q=|\bq|$.
(d), (h) $\langle u^x_{\bq}(\omega) u^x_{-\bq}(-\omega) \rangle$ for different persistence time $\tau\gamma$.
(a)-(d) are obtained with $K/(\gamma^2m)=10^{5}$, while (e)-(h) are obtained with $K/(\gamma^2m)=10^{3}$.
In all the panels, vertical dashed lines are used to denote the peak frequencies, i.e.\ $\Omega/\gamma$ and 
When it is not explicitly stated, the curves are obtained with $q=|\bq|=10^{-2}$, $\Omega/\gamma=30$, $ mv_0^2/T=10^2$, and $\tau\gamma=2$. 
}\label{fig:Fig1_dynamicalcorrelations}
\end{figure*}

From the explicit solution~\eqref{uqomegaequation}, the displacement correlation functions in the $(\bq,\omega)$ representation can be readily derived by utilizing the correlation properties of the noise. The detailed calculations are provided in Appendix \ref{spectralformcorrelations}, but here we present the expression for the dynamical correlations of the particle displacement with the same Cartesian index:
\begin{flalign}&
\langle  \tilde u^x_{\bq}(\omega) \tilde u^x_{-\bq}(\omega') \rangle=\frac{2\gamma}{(\omega^2-\omega_q^2 )^2 + \omega^2\gamma^2   }\times \nonumber\\
&\times\Bigl( \frac{T}{m }+v_0^2\tau\gamma \frac{(\omega^2+\Omega^2)\tau^2 +1}{ (1-(\omega^2  -\Omega^2) \tau^2)^2 +4\omega^2\tau^2}
\Bigr) \,2\pi\delta(\omega+\omega')  \, .
\label{uxuxomega}
\end{flalign}
%
%
For harmonic interactions, we observe an invariance for along $x$ and $y$, or in other words we have $\langle  \tilde u^y_{\bq}(\omega) \tilde u^y_{-\bq}(-\omega) \rangle=\langle  \tilde u^x_{\bq}(\omega) \tilde u^x_{-\bq}(-\omega) \rangle$.
The expression for $\langle  \tilde u^x_{\bq}(\omega) \tilde u^x_{-\bq}(-\omega) \rangle$ consists of two terms: i) a thermal contribution related to equilibrium phonons and ii) a non-equilibrium, active contribution with a different functional form.
i) corresponds with the first term within the parentheses in Eq.~\eqref{uxuxomega} and accounts for the displacement correlation stemming from thermal fluctuations in the presence of dissipative dynamics.  The amplitude of this equilibrium term is determined by the constant temperature $T$.
As is typical in equilibrium systems, the inertial term $1/\gamma$ governs a transition between monotonic decay and damped oscillatory behavior.
Specifically, for wavevectors satisfying $\omega_{\mathbf{q}}^2 > \gamma^2/4$, the poles of  $\mathcal{R}^{\hat{u}\hat{u}}_{\bq}(\omega)$ are complex (underdamped regime), while for $\omega_{\mathbf{q}}^2 < \gamma^2/4$ the poles become purely imaginary (overdamped regime).
This implies that in the underdamped regime dispersive phononic peaks at $\omega =\pm\omega_\bq$ are expected (Fig.~\ref{fig:Fig1_dynamicalcorrelations}~(a)-(d)). In the overdamped regime, however, a single peak at $\omega=0$ emerges (Fig.~\ref{fig:Fig1_dynamicalcorrelations}~(e)-(h)).  
The term ii) represents a non-equilibrium contribution arising from the active force with  its amplitude determined by the so-called active temperature, defined here as $v_0^2\tau\gamma$.
Although this term exhibits the same wave vector dependence as the thermal contribution, the non-equilibrium contribution is characterized by a prefactor that explicit depends on the frequency $\omega$ (see the second term in the parentheses in Eq.~\eqref{uxuxomega}).
Whether this frequency-dependent term can be interpreted as an effective temperature lies beyond the scope of this work.
While this frequency-dependent prefactor includes an explicit dependence on the persistence time $\tau$, as  noted previously~\cite{caprini2023entropons, caprini2023entropy}, we emphasize here that its value is significantly influenced by the chirality $\Omega$.
In the absence of chirality $\Omega$, the prefactor simplifies to $1/(1+\omega^2\tau^2)$, which is a function peaked at  $\omega=0$.
In contrast, when $\Omega>0$, the prefactor is peaked at the frequency $\omega=\Omega$.

In Fig.~\ref{fig:Fig1_dynamicalcorrelations}~(a) and~(e), the displacement correlations $\langle  \tilde u^x_{\bq}(\omega) \tilde u^x_{-\bq}(-\omega) \rangle$ are shown as a function of the frequency $\omega$ at fixed wave vector $\mathbf{q}$.
In the underdamped regime (Fig.~\ref{fig:Fig1_dynamicalcorrelations}~(a)), the dynamical correlations are characterized by four peaks. Two of these peaks occur at the phonon frequencies $\omega\pm \omega_\mathbf{q}$ and are driven by both the thermal and non-equilibrium contributions described in Eq.~\eqref{uxuxomega};
These peaks are only weakly influenced on the chirality $\Omega$ which is does not alter their positions and are not present in the overdamped regime (Fig.~\ref{fig:Fig1_dynamicalcorrelations}~(e)) where $\langle  \tilde u^x_{\bq}(\omega) \tilde u^x_{-\bq}(-\omega) \rangle$ has a single peak in the origin.
The two additional peaks, present in both the underdamped and overdamped regimes, are entirely induced by activity, particularly by chirality.
These peaks occur at frequencies $\omega=-\pm \Omega$ (Fig.~\ref{fig:Fig1_dynamicalcorrelations}~(b), (f)) and collapse to the origin as $\Omega\to0$.
Notably, these are non-dispersive peaks, as their positions do not depend on the wave vector $\mathbf{q}$
(Fig.~\ref{fig:Fig1_dynamicalcorrelations}~(c), (g)),
in contrast to the peaks at $\omega = \pm \omega_\mathbf{q}$, whose positions are determined by the relation $\omega_\mathbf{q}^2 \sim K\mathbf{q}^2$.
Finally, the persistence time $\tau$ does not influence the positions of the peaks but rather affects their widths and heights. Specifically, larger values of $\tau$ result in narrower and taller peaks 
 (Fig.~\ref{fig:Fig1_dynamicalcorrelations}~(d), (h)). 
 Consequently, we conclude that the persistence time has a dynamical effect analogous to inertia, while chirality can produce non-dispersive peaks at the frequency determined by the angular velocity $\Omega$.
 Since these effects and the influence of chirality arise in both overdamped and underdamped regimes, we will, in the following, evaluate static correlation functions exclusively in the overdamped case, where analytical results can be explicitly obtained.

\subsection{Spectral density of cross correlation and angular momentum}
One of the most distinctive traits of the chiral model is the existence of cross-correlations, meaning the $x$ component of displacement (or velocity) exhibits a non-zero correlation with its $y$ component. Moreover, swapping the components results in a change in the sign of the correlation:
$\langle  \tilde u^y_{\bq}(\omega) \tilde u^x_{-\bq}(-\omega) \rangle=-\langle  \tilde u^x_{\bq}(\omega) \tilde u^y_{-\bq}(-\omega) \rangle$.
Specifically, the cross $\omega$-correlations are purely imaginary, arising solely from active fluctuations and remaining unaffected by thermal noise:
\begin{flalign}\label{uxuyomega}
\langle  \tilde u^x_{\bq}(\omega) \tilde u^y_{-\bq}(\omega') \rangle=& i\,
 v_0^2\tau\gamma \frac{\gamma}{( \omega^2- \omega_{\bq}^2 )^2 + \omega^2\gamma^2   }\times\\
&\times  \frac{   4\omega\Omega \tau^2    }{   (1-(\omega^2-\Omega^2)\tau^2)^2 +4\omega^2\tau^2    } \, \,2\pi\delta(\omega+\omega') \nonumber\,,
\end{flalign}
where $i$ is the imaginary unit.
This expression is evaluated  in Appendix~\ref{spectralformcorrelations}.
The imaginary part of these cross-correlations qualitatively resembles the diagonal dynamical correlations of the displacement. Both types of correlations exhibit the same peaks, at $\omega=\pm \omega_{\mathbf{q}}^2$ and $\omega=\pm \Omega$ as well as the same parameter dependencies  (Fig.~\ref{fig:Fig1_dynamicalcorrelations}~(a)).
However, $\langle  \tilde u^x_{\bq}(\omega) \tilde u^y_{-\bq}(-\omega) \rangle$ is an odd function of $\omega$ and $\Omega$ 
and contains no thermal contributions. The amplitude of this correlation is solely determined by the active temperature
$v_0^2\gamma\tau$.
The presence of non-vanishing cross-correlations in particle displacement leads to a non-zero spectral density of angular momentum,  $\mathbf{M}$. 
This quantity is defined as the dynamical correlations between the cross-product of displacement and velocity. It serves as a measure of how the signal varies across a given frequency and wavevector, enabling the isolation of contributions from various length scales and timescales.
 By employing the solution~\eqref{uqomegaequation} and recalling that $\tilde{\mathbf{v}}_{\mathbf{q}}(\omega) = i \omega \tilde{\mathbf{u}}_{\mathbf{q}}(\omega)$, we derive (see Appendix~\ref{spectralformcorrelations}):
\begin{flalign}
\label{eq:spec_dens_M}
 {\tilde {\bf M}_\bq(\omega)}&=\frac{1}{2}\Bigl(\langle\tilde \bu_{\bq}(\omega)\times \tilde \vv_{-\bq}(\omega') \rangle+c.c\Bigr) \,2\pi\delta(\omega+\omega')\\
&=
 \frac{ 8  v_0^2\gamma^2}{( \omega^2- \omega_q^2 )^2 + \omega^2\gamma^2   }
 \frac{   \omega^2\Omega \tau^3  \,2\pi\delta(\omega+\omega')  }{ ( 1+\Omega^2\tau^2 -\omega^2 \tau^2 )^2 +4\omega^2 \tau^2 } \hat{\mathbf{z}} \nonumber\,,
\end{flalign}
where $\hat{\mathbf{z}}$ represents the unit vector along the vertical direction.
In this expression, the $\omega$-dependence of ${\tilde {\bf M}_\bq(\omega)}$
closely resembles that of the displacement-displacement correlation function~\eqref{uxuyomega}.
Specifically, ${\tilde {\bf M}_\bq(\omega)}$ and $\langle  \tilde u^x_{\bq}(\omega) \tilde u^y_{-\bq}(-\omega) \rangle$ share the same denominator, leading to identical imaginary or complex poles. The key distinction lies in the $\omega^2$ dependence in the numerator of ${\tilde{\bf M}_\bq(\omega)}$, which makes this function even with respect to $\omega$.
As a result, $ {\tilde {\bf M}_\bq(\omega)}$ exhibits two pairs of distinct peaks.
The first pair occurs very near the phonon frequencies, $\pm\omega_\bq$, while the second appears at a frequency close to the angular drift frequency, $\pm\Omega$.
Similar to the dynamical correlation of the displacement, the latter peak is non-dispersive, meaning that the frequency of its maximum is nearly independent of the wavevector $q$.
Furthermore, it is notable that the spectrum of the angular momentum remains independent of temperature, as it is driven solely by the active chiral noise.
We anticipate that the even $\omega$-dependence in Eq.~\eqref{eq:spec_dens_M} indicates that a chiral active crystal will exhibit a non-zero total angular momentum. This property arises solely due to chirality, as the total angular momentum vanishes for non-chiral active or passive crystals. This idea will be explored further in the next section through explicit real-space calculations.

\section{Spatial correlations}
\label{sec:staticcorrelations}
\label{Staticcorrelations}

In this section, we will focus on the static correlation functions in the steady state, i.e., for time  i.e.\ for time $t\to\infty$.
 In principle, time correlations can be obtained from their $\omega$-representation
  by analytically performing the necessary Fourier transforms through complex integration. This task becomes less demanding when considering the overdamped limit, where the propagator simplifies and becomes $\tilde G_{\bq}(\omega) =\Bigl(i \omega\gamma  +  \omega_{\bq}^2 \Bigr)^{-1}$.
Since chirality effects are observed in both the underdamped and overdamped regimes, we limit our discussion to the latter. This choice simplifies the analysis by significantly reducing the amount of analytical calculations required.

The equal-time correlations can be derived by integrating the dynamical correlations, such as Eq.~\eqref{uxuxomega} and Eq.~\eqref{uxuyomega}, over the entire frequency range  $\omega$. For example,
the displacement spatial correlations in the wave vector domain $\mathbf{q}$ at equal time are expressed as:
\begin{equation}
\langle  \hat u^\alpha_{\bq}(t) \hat u^\beta_{-\bq}(t) \rangle =\int_{-\infty}^\infty \frac{ d\omega}{2 \pi }
\int_{-\infty}^\infty \frac{ d\omega'}{2 \pi }\langle  \tilde u^\alpha_{\bq}(\omega) \tilde u^\beta_{-\bq}(\omega') \rangle 
\, 2\pi\delta(\omega+\omega') \,
\label{eq:def_corrqequaltime}
\end{equation}
where $\alpha,\beta=x,y$. 
Since the dynamical cross-correlations are odd functions of the frequency $\omega$, the equal-time correlations $\langle   \hat u^x_{\bq}(t)  \hat u^y_{-\bq}(t) \rangle=\langle   \hat u^y_{\bq}(t)  \hat u^x_{-\bq}(t) \rangle=0$ vanish in the steady state, as evident by considering the definition~\eqref{eq:def_corrqequaltime}.

This procedure can also be applied to other variables, such as velocity and self-propulsion to obtain general correlations of dynamical variables as a function of $\mathbf{q}$. 
By performing the inverse $\bq$-Fourier transform in two dimensions, one can derive the real space profile of these correlations:
\begin{flalign}
\langle   u^\alpha_{\mathbf{R}}(t)   u^\beta_{0}(t) \rangle = &
\frac{1}{N^2}\sum_{\bq=1}^N e^{-i \bq \cdot\mathbf{R}_{\bf{n}}} \langle \hat u^\alpha_{\mathbf{q}}(t) \hat u^\beta_{-\mathbf{q}}(t) \rangle \nonumber\\
&\approx  \frac{v_c }{(2\pi)^2}  \int_{BZ}  d \bq  \langle \hat u^{\alpha}_{\bq} \hat u^{\beta}_{-\bq}\rangle e^{-i \bq \cdot\mathbf{R}} \,,
\label{eq:def_spatialFourier}
\end{flalign}
where $\mathbf{R}$ is a real-space vector identifying the particle position and in the last approximation, the sum over $\mathbf{q}$ is replaced by an integral over the Brillouin region $BZ$. Here, $v_c$ represents the volume of the unit (Wigner-Seitz) cell of the crystal. 
By evaluating  Eq.~\eqref{eq:def_spatialFourier} at $\mathbf{R}=0$, we obtain the mean-square displacement.
    
For our convenience, rather than transforming the correlation functions from the
$\omega$ domain to the time domain, we have adopted a different but equivalent strategy, detailed in Appendix~\ref{timecorrelationsmethod}, 
which directly utilizes the time domain to evaluate the correlations. Using the methods outlined in Appendix ~\ref{timecorrelationsmethod}, the diagonal components of the equal-time displacement correlations are given by
\begin{eqnarray}
\langle   \hat u_{\bq}^x(t) \hat u_{-\bq}^x(t)\rangle =  \frac{T}{m}  \frac{1}{ \omega_{\bq}^2 }+
\frac{v_0^2 \tau\gamma}{ \omega_{\bq}^2 }
\frac{  1+\frac{\tau}{\gamma}\omega^2_{\bq}     }{   (1+\frac{\tau}{\gamma}\omega^2_{\bq})^2+\Omega^2 \tau^2     } \,.
\label{equaltimeuuv}
\end{eqnarray}
Similarly to the $\omega$-correlations, the expression for $\langle   u^x_{\bq}(t)  u^x_{-\bq}(t) \rangle$ 
contains two terms derived from direct integration of the two contributions in Eq.~\eqref{uxuxomega}.
The first term originates thermally and is proportional to the temperature of the thermal bath
, while the second term is entirely induced by the activity $(\propto \gamma \tau v_0^2)$ and vanishes in the equilibrium limit $v_0 \to 0$.  Consequently, the first term represents the standard passive contribution describing equilibrium dynamics, whereas the second introduces a novel source of fluctuations stemming from non-equilibrium effects.
Both the thermal and active terms contain a factor $1/\omega_{\bq}^{2}$ which diverges as $1/q^2$ as $\bq $ approaches $0$.

Except for the global $1/\omega_{\bq}^{2}$ factor, the passive term is independent of $\bq$, as expected in equilibrium.
In contrast, the active term contains an additional  $\bq$-dependence through $\omega_{\bq}^{2}$.
This extra dependence indicates the presence of non-equilibrium correlations in the displacement spectrum, which qualitatively differ from those at equilibrium. These displacement fluctuations, arising from non-equilibrium effects, are also present when chirality vanishes ($\Omega=0$)
and diminish in amplitude as $\Omega^2\tau^2$ increases.
The circular nature of the particles' trajectories reduces the spatial fluctuations of the particles and effectively increases the stiffness of their spring 
constants. The lack of long-range order in the two-dimensional crystal becomes evident when when $q$ satisfies the condition  $\tau<\gamma/\omega_q^2$.
This defines a critical value $q_{crit}$,  below which the modes make a significant contribution to the correlations. The critical value given by $\omega_{q_{crit}}^2=\frac{\gamma}{\tau}$.
This implies that the smallest wavevector, $q_{\text{min}}=2\pi/L$, determined by the system's size $L$, must be less than $q_{\text{crit}}$ to avoid finite size effects.

\begin{figure*}[!t]
\centering
\includegraphics[width=1\linewidth,keepaspectratio]{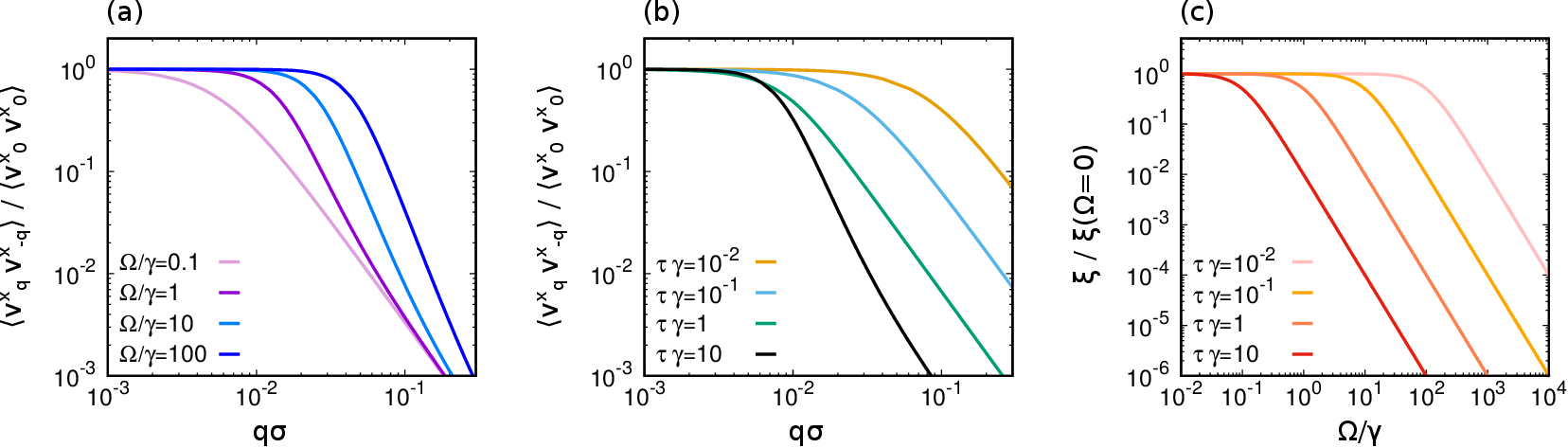}
\caption{
{\textbf{Chirality-suppressed spatial velocity correlations}}
(a)-(b) Spatial velocity correlations $\langle v^x_\bq v^x_{-\bq} \rangle$ (normalized by $\langle v^x_0 v^x_{-0} \rangle$) as a function of the wave vector modulus $q=|\bq|$ rescaled with the particle diameter $\sigma$. 
(a) shows $\langle v^x_\bq v^x_{-\bq} \rangle$ for different values of the reduced chirality $\Omega/\gamma$ at fixed reduced inertia $\tau\gamma=1$, while (b) shows $m(q)$ for different $\tau\gamma$ at fixed $\Omega/\gamma=1$.
(c) Correlation length $\xi$ of the spatial velocity correlations (normalized by correlation length calculated at zero chirality $\xi(\Omega=0)$) as a function of $\Omega/\gamma$ for different values of $\tau\gamma$. 
In all the panels, the curves are obtained with $ m v_0^2/T=10^2$ and $K/(\gamma^2m)=10^{3}$.
}\label{fig:Fig2_spatialvelocitycorrelations}
\end{figure*}

The mean square displacement of the particles from their equilibrium position is given by the expression:
$\langle \bu_\bn^2\rangle=     \frac{1}{N}\sum_{\bq} \langle\hat \bu_{\bq} \cdot \hat \bu_{-\bq}\rangle$, which 
upon converting the sums into integrals as in Eq.~\eqref{eq:def_spatialFourier}, reads:
\begin{equation}
\langle \bu_\bn^2\rangle=  \frac{v_c }{(2\pi)^2}  \int_{BZ}  d \bq  \langle \hat \bu_{\bq} \cdot \hat \bu_{-\bq}\rangle
\end{equation}
where  the integral extends to the first Brillouin zone (BZ). Utilizing Eq.~\eqref{equaltimeuuv} to perform the integral in polar coordinates, denoted as
($ \frac{v_c }{(2\pi)^2}  \int_{BZ}  d \bq \to \frac{v_c}{2\pi }  \,    \int_{q_{min}}^{q_D}    q dq $), with $q_{min}=2\pi/Na$ and  $q_D=\pi/a$.
Since the integral is peaked near $q=0$, we approximate the last fraction in Eq.~\eqref{equaltimeuuv} as $(1+\Omega^2\tau^2)^{-1}$ and use the small-$\mathbf{q}$ approximation for $\omega^2_{\bq} \approx 3K/(2m) q^2\sigma^2=c^2 q^2$ (the so-called linear Debye phonon spectrum), and obtain: 
\begin{equation}
\langle \bu_\bn^2\rangle
\approx \frac{1}{\pi}  \left[ \frac{v_c}{c^2}\Bigl(\frac{T}{m}+v_0^2 \frac{\tau\gamma }{1+\Omega^2\tau^2}\Bigr)\right]   \int_{q_{min}}^{q_D  } \frac{dq}{q}
\propto  \frac{\sigma^2}{\pi}    \ln N\, .
\label{approxun2}
\end{equation}
This logaritmic divergence for $N\to\infty$  aligns with the with the Mermin-Wagner theorem 
 predicting the absence of long-range translational order in a chiral active crystal.

However, the difference $\bu_\bn-\bu_\bn'$ between any pair of sites $(\bn$ and $\bn')$
 exhibits only finite fluctuations if their separation is finite, indicating that local order is preserved. 
Notably, in active chiral systems, both the mean square displacement and 
$(\bu_\bn-\bu_\bn')$ are reduced compared to their values in the achiral case. 
This reduction occurs because the trajectories in systems with chirality ($\Omega\neq 0$) 
are curved, and the positional fluctuations are reduced by activity. This behavior is consistent with the dynamics of an active particle in a harmonic potential \cite{caprini2023chiral}.
The particle number density can be expressed as the sum of delta functions positioned at $\rr_\bn(t)$, $\rho(\rr, t) = \sum_{\bn} \delta(\rr-\rr_\bn(t))$, where $\rr_\bn=\bR_\bn+\bu_\bn$. Its Fourier transform, assuming $\langle u_\bn\rangle$ to be independent of the site index, reads:
$
\langle \hat \rho_{\bq}\rangle=\frac{1}{N}\sum_\bn\langle e^{i  \bq\cdot \br_\bn}\rangle = \langle  e^{i  \bq\cdot \bu_\bn}\rangle\, f(\bq)
$ :
where
$
f(\bq)=\frac{1}{N}\sum_\bn e^{i \bq\cdot \bR_n}
$ represents the Fourier transformed density of the frozen lattice.
It is maximal when $\bq=G$, where $\bG$ is a reciprocal wave-vector. Since the displacements have a Gaussian distribution, we have:
\begin{equation}
 \langle e^{i  \bG\cdot \bu_\bn}\rangle= e^{-\langle (\bG\cdot\bu_\bn)^2\rangle/2}=e^{-W_\bG} \,.
 \end{equation}
Thus, the Fourier components of the density associated with the reciprocal wave-vectors are
 \begin{equation}
\langle \hat \rho_\bG\rangle=f(\bG) \, e^{-W_\bG}
\end{equation}
with
$
W_\bG=\frac{1}{2}\langle (\bG\cdot\bu_\bn)^2\rangle$.  Employing the approximate correlator and obtaining the following estimate:
\begin{flalign}
2W_\bG&=
\langle ({\bf G\cdot u}_n)^2\rangle\sim \frac{1}{4\pi} G^2 \left[ \frac{v_c}{c^2}\Bigl(\frac{T}{m}+v_0^2 \frac{\tau\gamma }{1+\Omega^2\tau^2}\Bigr)\right] \int_{\pi/L}^{q_D} dq \frac{1}{q}\nonumber\\
&\approx \frac{ G^2}{4\pi}\left[ \frac{v_c}{c^2}\Bigl(\frac{T}{m}+v_0^2 \frac{\tau\gamma }{1+\Omega^2\tau^2}\Bigr)\right]  \ln\left(\frac{L q_D}{ \pi} \right) \, .
\end{flalign} 
The Debye-Waller factor $e^{-W_\bG}$ vanishes for large systems as $L\to \infty$ ($N\to\infty$), and for any $\bG\neq 0$, $\langle \hat \rho_\bG\rangle=f(\bG) e^{-W_\bG} \to 0$. The displacement $\bu$ has infinite fluctuations, and the periodic order parameter, $\langle \hat \rho_\bG\rangle$, is washed out. Only the amplitude corresponding to $\bG=0$ remains finite in the infinite volume limit. However, unlike a liquid, the system displays power-law correlations. 

\subsection{Chirality effect on spatial velocity correlations}

\begin{figure*}[!t]
\centering
\includegraphics[width=1\linewidth,keepaspectratio]{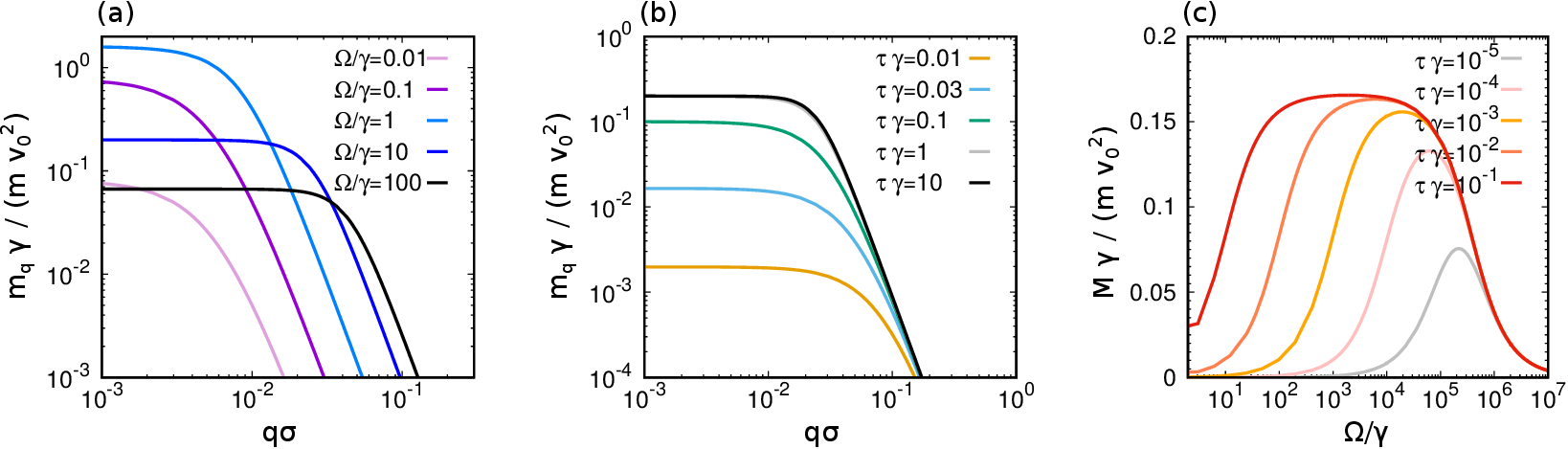}
\caption{
{\textbf{Chirality-induced angular momentum.}}
(a)-(b) Vertical component of the angular momentum $\mathbf{m}_\bq=\langle \mathbf{u}_\bq\times \mathbf{v}_{-\bq} \rangle +c.c=m_\bq \hat{\mathbf{z}}$ as a function of wave vector modulus $q=|\mathbf{q}|$ rescaled with the particle diameter. (a) shows $m_q$ for different values of the reduced chirality $\Omega/\gamma$ at fixed reduced inertia $\tau\gamma=1$, while (b) shows $m(q)$ for different $\tau\gamma$ at fixed $\Omega/\gamma=1$.
(c) Total angular momentum $M$ calculated by using Eq.~\eqref{magnlatticed} as a function of $\Omega/\gamma$ for different values of $\tau\gamma$. 
In all the panels, the curves are obtained with $m v_0^2/T=10^2$ and $K/(\gamma^2m)=10^{3}$.
}\label{fig:Fig3_angularmomentum}
\end{figure*}

An active crystal, composed of self-propelled particles without alignment interactions, exhibits spatial structures in velocity correlations. This phenomenon manifests in the formation of spatial domains where particles share the same (correlated) velocity, a behavior referred to as spontaneous velocity alignment~\cite{caprini2020spontaneous}.
This nonequilibrium collective effect has been observed in previous numerical studies, which predict an Ornstein-Zernike profile~\cite{caprini2020hidden, henkes2020dense, keta2022disordered, abbaspour2024long} with a correlation length analytically derived in terms of the model parameters~\cite{caprini2021spatial}. This correlation length increases with the square root of the persistence time and grows linearly with the spring strength.
Here, we evaluate the effect of chirality on spatial velocity correlations. The static velocity correlation can indeed be derived explicitly (see Appendix~\ref{timecorrelationsmethod}) and is given by:
\begin{equation}
\langle \hat v_{\bq}^\alpha(t) \hat v_{-\bq}^\alpha(t) \rangle
=\frac{T}{m}+v^2_0
\frac{ 1+\frac{\tau}{\gamma} \omega_q^2+\Omega^2\tau^2}{(1+ \frac{\tau}{\gamma}\omega_q^2)^2+\Omega^2\tau^2} \approx 
\frac{T}{m}+ \frac{v^2_0}{1+\frac{\tau}{\gamma}\frac{c^2 q^2}{1+\Omega^2\tau^2}}
\,.
\label{equaltimevv}
\end{equation}
where the latter approximation holds in the limit of small 
$\bq$.  Consequently, chirality does not modify the Ornstein-Zernike profile observed for non-chiral particles (Fig.~\ref{fig:Fig2_spatialvelocitycorrelations}(a)).
As with displacement correlations, the velocity correlations, $\langle v_{\bq}^\alpha(t) v_{-\bq}^\alpha(t) \rangle$,
 comprise two distinct contributions:
i) an equilibrium term, which remains constant ($\propto T$)
 and does not induce any spatial structure in the velocity field, and
ii) an active term, whose amplitude is governed by the swim velocity  $v_0^2$ 
 and vanishes in the equilibrium limit $v_0\to0$.
 The active term introduces a wave vector dependence, and  creates spatial correlations in the velocity field.

The role of chirality manifests in a faster decay, as shown at fixed  $\tau\gamma$
by varying the chirality parameter  $\Omega/\gamma$ (Fig.\ref{fig:Fig2_spatialvelocitycorrelations}~(a)),
and it decreases the correlation length $\xi$ (Fig.\ref{fig:Fig2_spatialvelocitycorrelations}~(c)) according to the formula:
\begin{equation}
\xi^2=\frac{3\tau }{4 \gamma} r^2 \Bigl( U''(\sigma)+ \frac{U'(\sigma)}{\sigma} \Bigr) \frac{1}{1+\Omega^2\tau^2} \,,
\label{coherencelength}
\end{equation}
where $U'$ and $U''$ denotes the first and second derivative of the potential evaluated at the lattice distance $\sigma$.
Equation~\eqref{coherencelength} reproduces previous results~\cite{caprini2020spontaneous, caprini2020hidden, henkes2020dense} in the limit of vanishing chirality, $\Omega=0$, and incorporates a chirality dependence consistent with the findings of Ref.~\cite{shee2024emergent}, which employed an alternative continuum approach.
The decrease in $\xi$ arises because chirality induces rotational motion in the particle trajectories, with a radius that decreases as $\Omega$ increases. This results in a smaller effective persistence length compared to the non-chiral case, leading to reduced spatial domains where velocities remain correlated.
In addition, consistent with the non-chiral case, a strong potential and/or a large value of $\tau$ (at fixed $\Omega$)
lead to slower decays of the spatial velocity correlations (Fig.\ref{fig:Fig2_spatialvelocitycorrelations}~(b)), 
 which corresponds to a larger correlation length $\xi$. By performing a Fourier transform of the profile in Eq.~\eqref{equaltimevv}, it is found that the velocities exhibit exponentially decaying correlations in real space. An estimate of the long-distance behavior of the velocity correlation function is:
 \begin{equation}
\langle \vv_\bR \cdot \vv_0\rangle\approx\frac{v_0^2}{(2\pi)^2}   \int_{BZ} d\bq \,\frac{e^{i\bq \cdot \bR}}{ (1+\xi^2\bq^2)} 
\approx   2v_0^2\frac{\xi^{-3/2}}{(8\pi R)^{1/2}}  e^{-  R/\xi}
\end{equation}
Here, we have omitted a delta-like contribution arising from the passive term, as  $T \ll v_0^2\gamma\tau$.
In the first approximation, we took the continuum limit, replacing the summation over discrete wavevectors with an integral over the first Brillouin zone. In the second approximation, we evaluated the correlation at long distances, deriving the asymptotic behavior of the spatial velocity correlations.

\subsection{Chirality-induced angular momentum}

Chirality primarily manifests through circular trajectories and a propensity for rotational motion. Consequently, it is not surprising that spatial velocity correlations are insufficient for effectively capturing novel collective effects induced by chirality.
To address this limitation, we propose using the angular momentum field~\cite{zhang2014angular} as a steady-state observable to quantify the system's handedness. This observable can be determined by integrating the spectral density of the angular momentum (Eq.~\eqref{eq:spec_dens_M}) over the frequency $\omega$, or directly from the equal-time off-diagonal correlation between velocity and displacement, yielding:
\begin{flalign}
\hat{ \mathbf{m}}_{\bq}(t)
&=
\frac{m}{2}(\langle \hat \bu_\bq(t)\times \hat \vv_{-\bq}(t) \rangle + c.c). = 
 \frac{2m\,v^2_0 \Omega\tau^2  }{(1+\frac{\tau}{\gamma}\omega_{\bq}^2)^2+\Omega^2\tau^2}\hat{\mathbf{z}} \nonumber\\
&\approx \frac{2m\,v^2_0 \Omega\tau^2  }{1+\frac{\tau}{\gamma}3 K\mathbf{q}^2+\Omega^2\tau^2} \hat{\mathbf{z}}\, .
\label{eq:angularmomentum_q}
\end{flalign}
Here, $\hat{\mathbf{z}}$ represents a unit vector in the vertical direction. In the final approximation, we have expanded $\omega_{\mathbf{q}}$ for small $\mathbf{q}$, retaining only the leading order in $\mathbf{q}$.
 Equation~\eqref{eq:angularmomentum_q} reveals that the angular momentum exhibits a $\mathbf{q}$-dependence, indicating the presence of a steady-state spatial structure in real space. Specifically, $\mathbf{m}_{\mathbf{q}}$ follows an Ornstein-Zernike profile with a correlation length of $2\xi$, where $\xi$ is defined in  Eq.~\eqref{coherencelength}.
 This behavior is illustrated for different values of $\Omega$ (Fig.\ref{fig:Fig3_angularmomentum}(a)) and different values of $\tau$ (Fig.~\ref{fig:Fig3_angularmomentum}(b)).
 Notably, the amplitude of $|\hat{\mathbf{m}}_{\mathbf{q}=0}|$ increases monotonically with $\tau$, whereas it exhibits a non-monotonic dependence on $\Omega$. This suggests that chirality may impart a non-monotonic effect on the total angular momentum $\mathbf{M}$. This global observable is obtained by summing contributions  $\mathbf{m}_{\mathbf{q}}$ from all modes $\mathbf{q}$ and normalizing by the number of particles  $N$:
 \begin{equation}
 {\bf M}
=\frac{m}{2N}\sum_{\bq} \Bigl(\langle\hat \bu_{\bq}\times\hat \vv_{-\bq} \rangle+c.c\Bigr)\,. 
\label{eq:M_global}
 \end{equation}
 The observable defined in Eq.~\eqref{eq:M_global} is a vector directed along the vertical direction and
 corresponds to the average angular momentum per particle because
 \begin{equation}
 {\bf M}=\frac{m}{N}\sum_{n=1}^N \left( \langle \bu_\bn\times \vv_\bn \rangle+\bR_\bn\times \langle\vv_\bn \rangle \right)=\frac{m}{N}\sum_{n=1}^N\langle\rr_\bn\times \vv_\bn \rangle\, ,
 \end{equation}
where the last equality holds because $ \langle\vv_\bn \rangle=0$ in the steady-state. 
As shown in Appendix \ref{Torquemomentum}, its modulus, $M$, is given by:
 \begin{eqnarray}&&
 M=   2 m v^2_0 \Omega\tau^2 
 \frac{1}{N}\sum_{\bq} 
 \frac{1 }{(1+\frac{\tau}{\gamma}\omega_q^2)^2+\Omega^2\tau^2}  \, ,
\label{magnlattice}
 \end{eqnarray}
and can be explicitly evaluated, by expanding $\omega_{\bq}\approx cq$ and approximating the sum in Eq.~\eqref{magnlattice} as an integral:
\begin{equation}
M\approx   \frac{v_c}{2\pi}  \frac{m \gamma v^2_0 }{c^2}  
    \arctan\Bigl( \frac{  \frac{ \frac{\tau}{\gamma} c^2 q_D^2  }{|\Omega| \tau }}
{1+ \frac{ 1+ \frac{\tau}{\gamma} c^2 q_D^2  }{\Omega^2 \tau^2 }}
  \Bigr) \,,
\label{magnlatticed}
 \end{equation} 
where $q_D$ is the Debye frequency chosen as a cutoff for the integral over $\mathbf{q}$. 
 For small $\Omega\tau$ the angular momentum grows linearly 
 and can be approximated as: 
\begin{equation}
M\approx    \frac{v_c}{2\pi} m v^2_0 \tau  \frac{1}{2\pi }
   \frac{\Omega\tau  q_D^2 }{ 1  +\frac{\tau}{\gamma} c^2 q_D^2  }\,.
\end{equation}
   In contrast, for large values of $\Omega\tau\gg1$, the angular momentum decreases 
   following: 
\begin{equation}
M\approx  \frac{v_c}{2\pi} m v^2_0 \tau \frac{1}{2\pi }\frac{ q_D^2  }{\Omega \tau }\,.
\end{equation}
   This results in a non-monotonic dependence on $\Omega$ 
which is evident for different values of $\tau$ 
as shown in Fig.~\ref{fig:Fig3_angularmomentum}(c).
This non-monotonicity can be understood by noting that for $\Omega=0$ (in the absence of chirality) the angular momentum vanishes due to symmetry.
Similarly, we expect $M$ to vanish again for $\Omega\to\infty$.
This behavior arises because chirality induces a bending of the trajectories of an active particle, thereby reducing displacement fluctuations compared to the non-chiral case, eventually suppressing them entirely for sufficiently large chirality.
This interpretation aligns with the findings from the study of a single chiral active particle in a harmonic potential \cite{caprini2023chiral}.
Let us note that, as demonstrated in Appendix \ref{Torquemomentum}, the torque, ${\bf T}$, 
exerted by the active chiral forces on the particles is
balanced by the torque exerted by the friction, the latter being proportional to the angular momentum 
multiplied  the friction coefficient, $\gamma$, yielding:
$|{\bf T}|=\gamma M$.

\subsection{Steady-state entropy production rate of a chiral solid }
\label{entropyproductionrate}

The entropy production serves as a measure of a system's irreversibility and its deviation from thermodynamic equilibrium~\cite{dabelow2019irreversibility, fodor2022irreversibility, o2022time}. In systems far from equilibrium, time-reversal symmetry is broken, and detailed balance no longer holds. As a result, the probability of a stochastic trajectory (forward path) differs from the probability of its time-reversed counterpart (backward path).
This asymmetry is quantified by the Kullback-Leibler divergence, expressed as the logarithm of the ratio between the probabilities, $P_f$, associated with forward paths, and $P_r$, associated with backward paths. This framework is used to define the entropy production rate, $\dot{S}$~\cite{seifert2012stochastic, fodor2016far, caprini2019entropy} as:
\begin{equation}
\dot{S} = \lim_{t\to\infty}\frac{1}{t}\left\langle\ln{\left(\frac{P_f}{P_r}\right)}\right\rangle \, ,
\label{eq:EPR}
\end{equation}
where a non-zero value of $\dot S$ differentiates non-equilibrium steady states from equilibrium states.
It is worth noting that the expression in \eqref{eq:EPR} is often referred to as the entropy production rate of the medium. However, this term coincides with the entropy production rate in the steady state, which is the primary focus of our interest.

In the case of non-chiral active systems, each chiral active particle is intrinsically far from equilibrium~\cite{speck2016stochastic,pietzonka2017entropy, pigolotti2017generic,caprini2019entropy, szamel2019stochastic, crosato2019irreversibility, chaki2018entropy, cocconi2020entropy} 
and on average, each equally contributes to the entropy production rate. The total entropy production rate per particle of a chiral active crystal can be expressed as the sum of two contributions 
\begin{equation}
\dot S=\dot S^v+\dot S^f \,.  
\label{eq:Sdot_sum}
\end{equation}
The first term $\dot S^v$ reads
\begin{equation}
\dot S^v=\frac{1}{N}\sum_{\bn} \langle {\bf f}^a_\bn(t) \cdot\vv_\bn(t) \rangle/T\,,
\end{equation}
and has the usual form of the entropy production rate of active particles, i.e.\ correspond to the work performed by the active force. The dependence on the chirality is implicitly contained in the equal-time correlations $ \langle {\bf f}^a_\bn(t) \cdot\vv_\bn(t) \rangle$. Approximating this correlations, leads to the following result (see Appendix~\ref{app:EPR}) 
\begin{flalign}
\dot S^v\approx &\frac{v_c}{4\pi} \frac{m\gamma^2 v^2_0  }{T} \frac{1}{  \tau c^2}
\Bigl\{ 
 \ln\Bigl(\frac{(1+\frac{\tau}{\gamma} c^2 q_D^2)^2+\Omega^2\tau^2 }{  1+\Omega^2\tau^2   }\Bigr)\nonumber\\
&+2 |\Omega| \tau  \arctan\Bigl( \frac{  \frac{ \frac{\tau}{\gamma} c^2 q_D^2  }{|\Omega| \tau }}
{1+ \frac{ 1+ \frac{\tau}{\gamma} c^2 q_D^2  }{\Omega^2 \tau^2 }}
  \Bigr)  
\Bigr\}\,.
\label{magnlatticed3b}
\end{flalign}
The entropy production rate contains an additional contribution arising from the active force dynamics, $\dot S^f$, and  given by the formula
\bea
\dot S^f=\frac{ \Omega \tau} {(m \gamma v_0)^2} \frac{1}{N}\sum_\bn \left\langle   {\bf f}^{a}_\bn  \times \dot {\bf f}^{a}_\bn   \right\rangle
= 2  \Omega^2\tau \,
\label{EPRchiralbath}
 \eea
which is  derived in Appendix \ref{app:EPR} using a path integral approach to calculate forward and backward trajectories.

The term $S^v$ (Eq.~\eqref{magnlatticed3b}), representing the work performed by the active force, is intuitively non-zero because both the velocity and the active force tend to align, regardless of whether the particles are chiral or harmonically confined. 
 $S^v$ consists of two contributions: i) 
a logarithmic term and ii) an arctan term.
i) has a functional form similar to that observed in non-chiral active particles, as discussed in Ref.~\cite{caprini2023entropy}. Thus, this term can be naturally interpreted as a translational contribution arising from the tendency of active particles to persistently moving in a typical direction. Consistently with our interpretaion, chirality reduces the magnitude of this term, which scales approximately as $\sim 1/\Omega^2$ for large chirality values.
By contrasts, the arctan contribution ii) emerges from the angular drift present in the dynamics of the active force: it disappears for vanishing chirality ($\Omega=0$) and was not present in previous formulations of the entropy production rate for crystals consisting of non-chiral active particles. This is a rotational contribution to entropy production which is indeed proportional to the chirality-induced angular momentum (Eq.~\eqref{magnlattice} and Eq.~\eqref{magnlatticed}) or, in other words, the torque exerted by the chiral active force.

From this, we can immediately conclude that the interplay between these terms results in an entropy production rate that increases with chirality $\Omega$. As $\Omega$ approaches zero, $\dot S^v$ tends to a finite limit, which corresponds to the entropy production rate of a solid composed of achiral AOUP particles. Conversely, $\dot S^v$ saturates for large values of $\Omega$, while $\dot S^f$ vanishes at $\Omega=0$ and diverges quadratically as $\Omega$ approaches infinity.

\section{Time-dependent correlations }
\label{timedependent}

In addition to decaying with particle separation, correlations also diminish over time. Understanding the time-dependent behavior of correlations is essential for unraveling the range and duration of collective effects in active matter systems.

As in the previous section, we will focus on the overdamped regime, where $\gamma^2/4 \gg \omega_{\bq}^2$,
to facilitate analytical progress and assess the impact of chirality. 
The temporal correlations, such as $\langle   \hat u^\alpha_{\bq}(t)  \hat u^\beta_{-\bq}(0) \rangle $ are obtained in the wave vector domain $\mathbf{q}$.
 Temporal autocorrelations for the same particle coordinates are calculated by summing  over $\mathbf{q}$ (or integrating over the Brillouin zone $BZ$):
\begin{equation}
\langle   u^\alpha_{0}(t)  u^\beta_{0}(t) \rangle = 
\frac{1}{N}\sum_{\bq} \langle\hat  u^\alpha_{\mathbf{q}}(t)  \hat u^\beta_{-\mathbf{q}}(t) \rangle \, \approx  \frac{v_c }{(2\pi)^2}  \int_{BZ}  d \bq  \langle \hat u^{\alpha}_{\bq}  \hat u^{\beta}_{-\bq}\rangle  \,
\label{eq:def_temporalcorr_time}
\end{equation}

\subsection{Time-dependent displacement and velocity correlations in the overdamped Regime}

After lengthy but straightforward calculations reported in Appendix \ref{timecorrelationsmethod} (see Eqs.\eqref{eqC8} and \eqref{eqC9}), we obtain the following formula for the displacement correlations with $t\geq 0$:
  \begin{flalign}
\langle  \hat u_{\bq}^x(t) \hat u_{-\bq}^x(0)\rangle=& \frac{T}{m}\frac{1}{\omega^2_{\bq}}  e^{-\frac{\omega^2_{\bq}}{\gamma} t}+\\ 
&+ \frac{v_0^2 }{\omega^2_{\bq}}  \frac{\tau\gamma}{\left[1-\frac{\tau}{\gamma}\omega^2_{\bq}\right]^2+\Omega^2 \tau^2 }    \frac{  \mathcal{F}^{xx}_\bq(t)  }{   \left(1+\frac{\tau}{\gamma}\omega^2_{\bq}\right)^2+\Omega^2 \tau^2     } \nonumber
\label{diagonaluu}
\end{flalign}
where
\begin{flalign}
\mathcal{F}^{xx}_\bq(t)=&
 \left(1-\left(\frac{\tau}{\gamma}\omega^2_{\bq}\right)^2+\Omega^2\tau^2\right)  e^{-\frac{\omega^2_{\bq}}{\gamma} t} + 2\frac{\tau}{\gamma}\omega^2_{\bq} \,\Omega\tau e^{-\frac{t}{\tau}}  \sin(\Omega t) \nonumber\\
&- \frac{\tau}{\gamma}\omega^2_{\bq} \,	\left(1-\left(\frac{\tau}{\gamma}\omega^2_{\bq}\right)^2-\Omega^2\tau^2\right)  e^{-t/\tau} \cos(\Omega t) \,.
\end{flalign}
By symmetry, we have $ \langle   u_{\bq}^y(t) u_{-\bq}^y(0)\rangle =\langle   u_{\bq}^x(t) u_{-\bq}^x(0)\rangle$.
The first term in Eq.~\eqref{diagonaluu} represents the contribution of  thermal noise, while the second and third terms are due to the active force and have a non-equilibrium origin.
Chirality reduces the amplitude of activity-induced fluctuations compared to the non-chiral case, leading to damped oscillatory behavior with frequency $\Omega$, as evident from the presence of $\sin(\Omega t)$ and $\cos(\Omega t)$.
We classify fluctuations based on their wavevector, $\bq$: i) Fluctuations with persistence times exceeding the viscous time associated with the mode of wavevector $\bq$ ($\tau\gg\gamma/\omega_q^2$). ii) Long-wavelength fluctuations for which $\tau \ll \gamma/\omega_q^2$. In the first case, the slowest relaxation modes are represented by terms proportional to $ e^{-\frac{1}{\tau}  t}$. For time separations $t$ exceeding $\tau$, the most relevant terms in the correlations vary exponentially as $e^{- \frac{\omega^2_q}{\gamma} t}$, with slower decay for smaller wavevectors $\bq$. After a short initial transient, the amplitude of these long-wavelength modes contributes significantly to real-space correlations.

 Using Eq.~\eqref{eqC10}, we obtain the time correlation between diferent spatial components of displacement:
\begin{flalign}&
\langle  \hat u_{\bq}^x(t) \hat u_{-\bq}^y(0)\rangle=- \frac{v_0^2\tau^2}{\left[1-\frac{\tau}{\gamma}\omega^2_{\bq}\right]^2+\Omega^2 \tau^2 }  \frac{  \mathcal{F}^{xy}_\bq(t)  }{   \left(1+\frac{\tau}{\gamma}\omega^2_{\bq}\right)^2+\Omega^2 \tau^2     }
 \label{offdiagonaluu}
\end{flalign}
where
\begin{flalign}
\mathcal{F}^{xy}_\bq(t)= &2\Omega\tau \left( e^{-\frac{\omega^2_{\bq}}{\gamma} t}- e^{-t/\tau} \cos(\Omega t)  \right) \nonumber \\
&-\,\left(1-\left(\frac{\tau}{\gamma}\omega^2_{\bq}\right)^2-\Omega^2\tau^2\right) e^{-t/\tau} \sin(\Omega t)  \,.
 \label{offdiagonaluu}
\end{flalign}
Exchanging the components ($xy\to yx$) changes the sign of the correlation, i.e., $\langle \hat u_{\bq}^y(t) \hat u_{-\bq}^x(0)\rangle = -\langle \hat u_{\bq}^x(t) \hat u_{-\bq}^y(0)\rangle $, resulting in vanishing equal-time off-diagonal elements. In the short-time regime, the correlation varies linearly with $t$, contrasting with the diagonal correlation.
Asymptotically ($t\gg\tau$), the leading term of the two-time correlation is proportional to $\Omega e^{-\omega_{\bq}^2 t/\gamma}$, with the prefactor being an odd function of $\Omega$.

By using the method of Appendix \ref{timecorrelationsmethod}, we obtain the two-time diagonal elements of the velocity correlations, which can be expressed as follows:
%
 \begin{flalign}
&\langle  \hat v_\bq^x(t) \hat v_{-\bq}^x(0)\rangle= \frac{T}{m} e^{-\frac{\omega^2_{\bq}}{\gamma} t} -v_0^2 \tau\gamma \frac{\mathcal{R}_{\bq}(t)}{[1-\frac{\tau}{\gamma}\omega^2_\bq]^2+\Omega^2 \tau^2 } 
\label{vxvxtimecorrel}
 \end{flalign}
where
\begin{flalign}
   \mathcal{R}_\bq(t)&= \frac{\omega^2_\bq}{\gamma^2}  e^{-\frac{\omega^2_\bq}{\gamma} t} \bigl[ 1- 2\frac{\tau}{\gamma}  \omega^2_\bq \,\frac{  (1+\frac{\tau}{\gamma}  \omega^2_\bq)  }{   (1+\frac{\tau}{\gamma}  \omega^2_\bq)^2+\Omega^2 \tau^2     }  \bigr]  +2\frac{\Omega}{\gamma}  e^{-t/\tau}  \sin(\Omega t)  \nonumber\\
&
  + \frac{\tau}{\gamma} (\frac{1}{\tau^2}-\Omega^2)   e^{-t/\tau} \cos(\Omega t)  
- 2\frac{\tau}{\gamma} \frac{  1  }{   (1+\frac{\tau}{\gamma}\omega^2_\bq)^2+\Omega^2 \tau^2     }  \times\nonumber\\
&\times\Bigr\{(1+\frac{\tau}{\gamma}\omega^2_\bq)   \bigl[(\frac{1}{\tau^2}-\Omega^2)  e^{-t/\tau}\cos(\Omega t) +2\frac{\Omega}{\tau}  e^{-t/\tau}  \sin(\Omega t) \bigr]        \nonumber\\
&-\Omega\tau  (\frac{1}{\tau^2}-\Omega^2)     e^{-t/\tau}  \sin(\Omega t)  +2 \Omega^2   e^{-t/\tau}  \cos(\Omega t)    \Bigl\} \,.
\end{flalign}
 Again, we can identify a thermal term proportional to $T$ that decays with a characteristic time $\gamma/\omega^2_{\bq}$ and a non-equilibrium term whose amplitude is given by the active temperature $v_0^2\tau\gamma$. This second term is characterized by an exponential decay governed by two typical timescales: $\gamma/\omega^2_{\bq}$ (as in the passive term) and the persistence time $\tau$. As with displacement correlations, the chirality $\Omega$ induces temporal oscillations for a short duration.
 
 By differentiating  Eq.~\eqref{offdiagonaluu}, we derive the temporal velocity cross-correlations $\langle \hat{v}{\mathbf{q}}^x(t) \hat{v}{-\mathbf{q}}^y(0) \rangle$. Similar to the displacement in Eq.~\eqref{offdiagonaluu}, these correlations are antisymmetric with respect to the exchange of indices $x$ and $y$, and vanish both at $t = 0$ and as $t \to \infty$.
Here, we simply present the expression in the long-time regime where $\omega_{\mathbf{q}}^2 \tau / \gamma \ll 1$:
\begin{flalign}
\langle  \hat v_\bq^x(t) \hat v_{-\bq}^y(0)\rangle \approx&
- 2 \Omega\tau v^2_0  \frac{1}{\left(1-\frac{\tau}{\gamma}\omega^2_q \right)^2+\Omega^2 \tau^2} \times\nonumber\\
&\times\frac{1}{  (1+  \frac{\tau}{\gamma} \omega^2_q)^2+\Omega^2\tau^2}  \left(\frac{\tau}{\gamma} \omega^2_q\right)^2 e^{-\frac{\omega^2_q}{\gamma}t} \, .
\label{eq:approx_vxvy_time}
 \end{flalign}
In the long-time regime, chirality contributes to these temporal correlations by decreasing their amplitude. This result is consistent with the reduction of persistence length for $\Omega>0$.
Conversely, chirality induces circular trajectories, which manifest as short-time oscillations in the temporal correlations. These oscillations are neglected in the approximation of Eq.~\eqref{eq:approx_vxvy_time}.

\subsection{Time-dependent Density Self-correlation Function and mean-square-displacement}

\begin{figure*}[!t]
\centering
\includegraphics[width=1\linewidth,keepaspectratio]{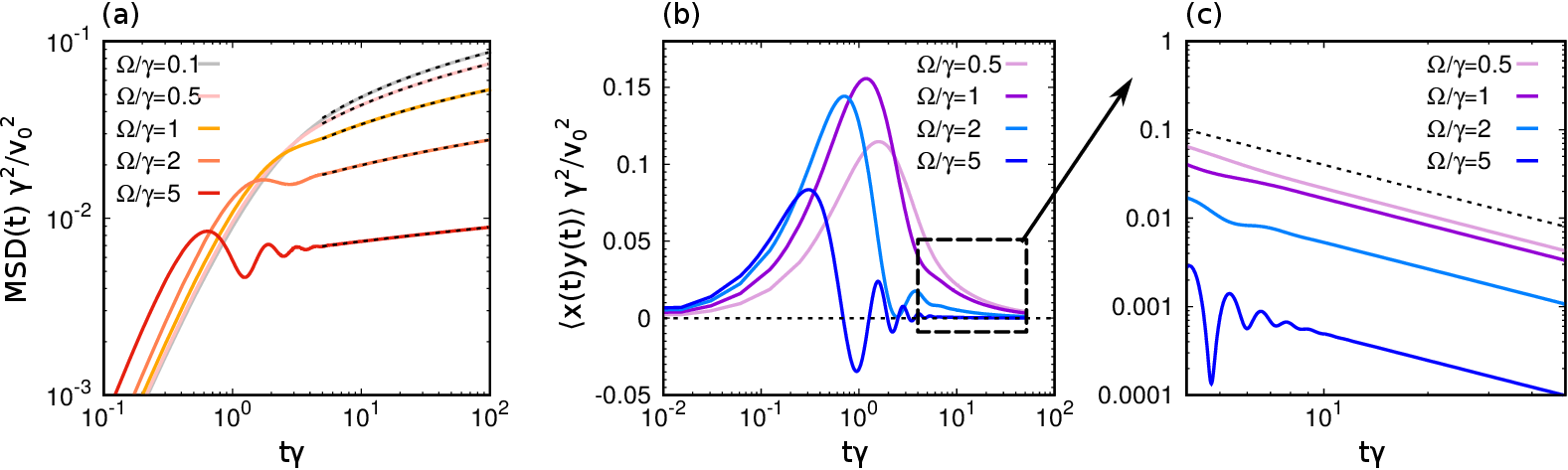}
\caption{
{\textbf{Time-dependent properties.}}
(a) mean-square-displacement $\text{MSD}(t)=\langle(\bu_\bn(t)-\bu_\bn(0))^2 \rangle$ as a function of time $t$ for different values of the chirality $\Omega /\gamma$ normalized by the inertial time. Colored lines are obtained by integrating Eq.~\eqref{sigma0b1} over $\mathbf{q}$ while black dashed lines are realized by fitting a logarithm function $a\log{(t b)}$ where $a$ and $b$ are fitting parameters. These fits confirm the approximation~\eqref{sigmaanalit}.
(b)-(c) temporal cross correlation $\langle u^x_\bn( t)u^y_\bn(0)\rangle$ as a function of time $t$ for different values of $\Omega/\gamma$. Colored lines are obtained by integrating Eq.~\eqref{offdiagonaluu} over $\mathbf{q}$.
(c) is a zoom on the square area in panel (b), where the $1/t$ scaling is shown as a black dashed line.
In all the panels, the curves are obtained with $m  v_0^2/T=10^2$ and $K/(\gamma^2m)=10^{3}$ and $\tau \gamma=1$.
}\label{fig:Fig4_timecorr}
\end{figure*}

Although the harmonic lattice is a linear model, deriving an analytical expression for the self-density-density correlation is challenging due to its nonlinear dependence on the displacement correlation. We consider the average $\left\langle \rho_n(\mathbf{x},t)  \rho_n(\mathbf{x}',0)\right\rangle$.
The expression $\rho_\bn(\bx,t)\equiv \delta^{(2)}(\bx-\rr_\bn(t) )$ represents the probability density that the $\bn$-th particle is located at time $t$ within a small volume centered around the point $\bx$.

Given the translational invariance of the system, we integrate over the entire volume and average over the distribution of particle positions. This leads us to define ${\cal G}_{inc}(\br,t)$ as follows:
${\cal G}_{inc}(\br,t)=\left\langle \delta^{(2)}(\rr_\bn(0)+\br-\rr_\bn(t) ) \right\rangle$.
Using the displacement variables and the Fourier representation of the Dirac delta function we have:
\begin{flalign}
{\cal G}_{inc}(\rr,t)
&=\frac{1}{(2\pi)^2}  \int d^2{\bf p} \,e^{ -\frac{p^2}{2} \bigl\langle(\bu_\bn(t)-\bu_\bn(0))^2\bigr\rangle} \, e^{i  {\bf p}\cdot \rr} \nonumber\\ &=\frac{1}{2 \pi } \frac{1}{\text{MSD}(t)} e^{-\frac{1}{2} \rr^2/\text{MSD}(t)}
\label{diffusionequation}
\end{flalign}
where $\text{MSD}(t)=\langle(\bu_\bn(t)-\bu_\bn(0))^2 \rangle$.
In the Gaussian expression above, the dependence on the chiral parameter $\Omega$ is embedded within the mean square displacement, $\text{MSD}(t)$, which can be computed by using the Green-Kubo relation, i.e.\ by double integrating over time the autocorrelation of the velocity~\eqref{vxvxtimecorrel} and then performing the integration over the wave vector $\mathbf{q}$.
This leads to the following formula:
 \begin{flalign}&
\text{MSD}(t)=\frac{1}{2\pi^2}\int_{BZ} d\bq\,\Bigl\{\frac{T}{m} \frac{1}{\omega^2_{\bq}}  [1-e^{-\frac{\omega^2_{\bq}}{\gamma} t}] \\
&+\frac{ v_0^2 \tau\gamma}{[1-\frac{\tau}{\gamma}\omega^2_{\bq}]^2+\Omega^2 \tau^2 }\Bigl[   \mathcal{H}_\bq(t) +  \mathcal{W}_\bq(t) 
+\Omega\tau e^{-t/\tau}  \sin(\Omega t)\Bigr ]\Bigr\} \nonumber\,.
\label{sigma0b1}
 \end{flalign}
where
\begin{flalign}
& \mathcal{H}_\bq(t) =\frac{1}{\omega^2_{\bq}}  [1-e^{-\frac{\omega^2_{\bq}}{\gamma} t}] + \frac{\tau}{\gamma}   [ 1-e^{-t/\tau} \cos(\Omega t)] \\
& \mathcal{W}_\bq(t) =
- 2\frac{\tau}{\gamma} \frac{  1+\frac{\tau}{\gamma}\omega^2_{\bq}  }{   (1+\frac{\tau}{\gamma}\omega^2_{\bq})^2+\Omega^2 \tau^2     }     \bigl[2-e^{-t/\tau}\cos(\Omega t)-e^{-\frac{\omega^2_{\bq}}{\gamma} t}\bigr]\,.
\end{flalign}
This observable provides insight into the particle's ability to explore the space and is illustrated in 
 Fig.~\ref{fig:Fig4_timecorr}~(a) 
 through the direct integration of formula~\eqref{sigma0b1} for various values of chirality,
  $\Omega$, using an upper cutoff $q_D$. 
Notably, the $\text{MSD}(t)$ 
displays a quadratic growth with respect to $t$ in the short-time regime
 $t < \tau$:
 \begin{eqnarray}&&
  \text{MSD}(t)\approx t^2\,      \frac{v_c}{4\pi^2} \int d\bq\, 
 \langle \hat \vv_{\bq}(0) \hat \vv_{-\bq}(0) \rangle \,.
  \label{selfvelocity0}
\end{eqnarray}
The variance $\langle\hat \vv_{\bq}(0) \hat \vv_{-\bq}(0) \rangle$ is given by Eq.~\eqref{vxvxtimecorrel} at $t=0$ and depends on the swim velocity and chirality.
This $t^2$ behavior is typical of active matter systems and arises from the self-propulsion force, which induces ballistic behavior over short time scales.
For sufficiently large chirality compared to the persistence time, such that $\Omega\tau >1$, the mean squared displacement, $\text{MSD}(t)$, displays oscillations which become more pronounced as $\Omega$ increases (Fig.~\ref{fig:Fig4_timecorr}~(a)). These oscillations emerge for intermediate times and are suppressed in the long-time regime, where the dominant contribution to the integral comes from the small wavevector region.
Consequently, the long-time behavior can be approximated as follows:
  \begin{equation}
  \text{MSD}(t)\approx  \frac{v_c}{\pi} \frac{2m}{3K \sigma^2}\Bigl( \frac{T}{m}+\frac{  v^2_0 \tau\gamma }{1+\Omega^2\tau^2 } 
\Bigr) \ln t \,.
\label{sigmaanalit}
   \end{equation}
The $\text{MSD}(t)$ exhibits a logarithmic divergence with respect to $t$,  similar to the passive case.
Indeed, expression~\eqref{sigmaanalit} contains a passive contribution proportional to $T$ and an active contribution proportional to the active temperature $v_0^2\tau\gamma$. The latter is reduced by chirality as evident in Fig.~\ref{fig:Fig4_timecorr}~(a).
Additionally, chirality induces a temporal profile in the cross-correlations, $\langle u_{\bn}^x( t) u_{\bn}^y(0)\rangle$, obtained by numerically integrating Eq.~\eqref{offdiagonaluu} over $\bq$ using the continuum approximation~\eqref{eq:def_temporalcorr_time}. Fig. \ref{fig:Fig4_timecorr} (b) displays this observable as a function of time for various values of $\Omega$.
Rather than presenting the lengthy expressions, we will concentrate on the short-time and long-time regimes. The short-time regime exhibits linear growth, which can be approximated by:
\begin{equation}
|\langle u^x_\bn( t)u^y_\bn(0)\rangle| = v_0^2 \tau \frac{ \Omega \tau}{(1+\frac{\tau}{\gamma} \omega_{\bq}^2)^2+\Omega^2\tau^2} \, t \,.
\end{equation}
Conversely, in the long-time regime, employing the same methodology as in deriving Eq.~\eqref{sigmaanalit}, yields the expression:
\begin{eqnarray}
| \langle u^x_\bn( t)u^y_\bn(0)\rangle|
\approx
 v_0^2 \frac{2m}{3K \sigma^2} \frac{\gamma\Omega\tau^3}{(1+\Omega^2 \tau^2)^2} \frac{1}{ t} \,.
\end{eqnarray}
Thus, for large $t$, temporal cross-correlations have a long-range $\sim 1/t$ decrease.
In agreement with the previous observations in the frequency and wave vector representation, the coefficients of the cross-correlations are non-monotonic functions of the chirality $\Omega$.

\section{Summary and conclusions}\label{sec:conclusions}
In this paper, we studied a two-dimensional harmonic solid composed of chiral active particles, which are modeled by incorporating an angular drift into the active force term.
We analyzed the displacement and velocity correlation functions in both the time and frequency domains. A harmonic chiral active crystal adheres to the Mermin-Wagner theorem, exhibiting a diverging displacement-displacement correlation in two dimensions. Furthermore, as in the non-chiral case, these chiral crystals display spatial velocity correlations. While chirality does not significantly alter the functional form of these correlations, it reduces their correlation length. This result is consistent with the continuum theoretical approach and the numerical results obtained in Ref.~\cite{shee2024emergent}.

The distinctive properties of chiral crystals are highlighted by a non-vanishing angular momentum, defined as the vector product of the displacement and velocity of the particles. The angular momentum stems from the torque exerted by the active force, which has a preferential direction due to chirality. The angular momentum, which is zero for a crystal consisting of non-chiral active particles,
shows a non-monotonic dependence on chirality and possesses a spatial structure.

Chirality also influences the dynamical properties of the crystal by altering the spectrum of particle displacements, introducing a non-dispersive peak at the chiral frequency. This peak appears in both the underdamped and overdamped regimes, corresponding to the observation of a phononic peak or a peak at vanishing frequency, respectively.
Moreover, the unique effects of chirality manifest in non-zero cross-correlations between different Cartesian components of displacement-displacement and velocity-velocity dynamical correlations in the frequency domain.
The presence of this peak leads to temporal oscillations in the autocorrelations of particle displacement and mean-square displacement, which we have analytically predicted.
Finally, chirality contributes an additional entropy production term, specifically associated with angular momentum.

Beyond the numerical investigation of a densely packed assembly of soft spheres~\cite{caprini2021spatial,caprini2020active} driven by chiral forces, our predictions can be tested in experimental active matter systems exhibiting chirality.
Promising candidates include systems of densely packed chiral active colloids~\cite{massana2021arrested} or colloids subjected to a magnetic field, as well as macroscopic systems of active granular particles~\cite{kumar2014flocking, briand2018spontaneously, caprini2024dynamical, siebers2023exploiting}. In these macroscopic setups, particles can be connected using rigid rods~\cite{caprini2024spontaneous} or springs~\cite{baconnier2022selective}, while chirality can be introduced by modifying the particle shape~\cite{scholz2018rotating, vega2022diffusive, scholz2021surfactants}.

Lastly, our exactly solvable model may bridge the gap between macroscopic theories of chiral crystals with odd elastic properties~\cite{fruchart2022odd}. These macroscopic elastodynamic theories rely on an antisymmetric elastic matrix. By employing coarse-graining techniques, one can derive the effective elastic matrix for our model, providing insights into the relationship between chirality and odd elasticity.

\appendix

\begin{strip}

\section*{Appendices}

\section{Definition of time and space Fourier transforms}

\label{definitionsfouriertransform}

Fourier transforms are performed in the time and space domains. Even if the explicit dependence on frequency $\omega$ (time $t$) or wave vector $\bq$ (particle index $\bn$) is always explicit, we adopt the following notation: i) a vertical bar over the variable denotes time Fourier transform in the domain ($\omega$, $\bn$). ii) The hat is used for the spatial Fourier transform in the domain ($t$, $\bq$). iii) The tilde is employed for the time and space Fourier transforms in the domain ($\omega$, $\bq$).
Specifically, we define the time Fourier transform of the particle displacement, which is denoted by the tilde symbol and an explicit dependence on the frequency $\omega$, as
\begin{equation}
  \bar{\mathbf{u}}_\bn(\omega)=\int_{-\infty}^\infty dt e^{i\omega t}
 \mathbf{u}_\bn(t) 
\label{eq:FourierTransform}
\end{equation}
while the inverse time Fourier transform reads
\begin{equation}
\mathbf{u}_\bn(t)= \int_{-\infty}^\infty \frac{d\omega}{2\pi} e^{-i\omega t}
\bar{\mathbf{u}}_\bn(\omega)   \,.
\end{equation}
Similar definitions hold for the other variables, such as velocity, active force and noise variables. 
Specifically, delta-correlated white noises, such that $\langle \xi_n(t) \xi_m(t') \rangle=\delta(t-t')  \delta_{nm}$,
satisfy the following relation in Fourier space
\begin{equation}
\langle \bar \xi_n(\omega) \bar \xi_m(\omega') \rangle=\int_{-\infty}^\infty dt e^{i\omega t} \int_{-\infty}^\infty dt' e^{i\omega t'}
\langle \xi_n(t) \xi_m(t') \rangle=\int_{-\infty}^\infty dt e^{i\omega t} \int_{-\infty}^\infty dt' e^{i\omega t'}\, \delta(t-t')  \delta_{nm}=2\pi\,\delta_{nm} \,\delta(\omega+\omega') \,,
\end{equation}
and
$\int_{-\infty}^\infty dt e^{i\omega t}
\langle \xi_n(t) \xi_m(0) \rangle=   \delta_{nm}$.
In a similar way, we can define the discrete spatial Fourier transform of the displacemente vector as
\begin{equation}
\hat{\mathbf{u}}_\bq(t)=\frac{1}{\sqrt N}\sum_{\bn}  e^{i \bq \cdot \bR_\bn} \mathbf{u}_\bn(t) \,,
\label{eq:spatial_FourierTransform}
\end{equation}
where $\mathbf{R}_\bn$ is a vector identifying the position of each lattice site and $\bq=(q_x, q_y)$ is a discrete wave vector and $N$ is the total number of particles.
The inverse discrete spatial Fourier transform is defined as
\begin{equation}
\mathbf{u}_\bn(t)=\frac{1}{\sqrt N}\sum_{\bn}  e^{-i \bq \cdot \bR_\bn}\hat{ \mathbf{u}}_\bq(t) \,.
\end{equation}
The combination of the time and space Fourier transforms leads to the definition~\eqref{eq:FTdef}.


\vskip20pt
\section{Derivation of the dynamical correlations}
\label{spectralformcorrelations}

In this Appendix, we derive the dynamical correlation of the particle displacement in the frequency $\omega$ and wave vector $\bq$ domains, i.e.\ Eqs.\eqref{uxuxomega}, \eqref{uxuyomega}
As a first step to finding analytical solutions, we consider the Fourier transforms in time and space of the dynamics~\eqref{eq:activeforce} for the active force and the dynamics~\eqref{dynamicequation2} for the particle displacement.
Specifically, the Fourier transform of the active force dynamics expressed in Cartesian components reads:
\begin{subequations}
\begin{eqnarray}&&
i\omega \tilde f^{a,x}_\bq(\omega)
+\frac{1}{\tau} \tilde f^{a,x}_\bq(\omega) +\Omega \tilde f^{a,y}_\bq(\omega)
= m\gamma v_0 \sqrt{ \frac{2 }{\tau}} \tilde \chi^x_\bq(\omega)
\label{eqfax}\\&&
i\omega \tilde f^{a,y}_\bq(\omega)
+\frac{1}{\tau} \tilde f^{a,y}_\bq(\omega) -\Omega \tilde f^{a,x}_\bq(\omega)
= m\gamma v_0 \sqrt{ \frac{2 }{\tau}} \tilde \chi^y_\bq(\omega)
\label{eqfay}
\end{eqnarray}
\end{subequations}
where we have used superscript to denote the $x$ or $y$ Cartesian component.
By contrast, the Fourier transform of the displacement dynamics reads
\bea&&
\Bigl(-\omega^2+i \omega\gamma  
+  \omega_{\bq}^2 \Bigr) \tilde \bu_\bq(\omega)= \frac{\tilde {\bf f}^a_\bq(\omega)}{m} + \sqrt{2\gamma \frac{T}{m }} \,\tilde{\xxi_\bq}(\omega) \,,
\label{eq:dyn_disp_app}
\eea
which corresponds to Eq.~\eqref{uqomegaequation} upon defining $\tilde{\boldsymbol{\mathcal{R}}}_\bq^{\hat{u}\hat{u}}(\omega)= \hat{G}_\bq(\omega) \mathbf{I}$ via the propagator $\hat{G}_\bq(\omega)$, introduced in Eq.~\eqref{underdampG}.
We remark that the Fourier transforms of the white noises satisfy the following properties:
\bea&&
\label{eq:noise_chi}
\langle \tilde \chi^\alpha_\bq(\omega) \tilde \chi^\beta_{-\bq}(\omega') \rangle=2\pi \,\delta^{\alpha\beta}\,\delta(\omega+\omega')
\\&&
\langle \tilde \xi^\alpha_\bq(\omega) \tilde\xi^\beta_{-\bq}(\omega') \rangle=2\pi\,\delta^{\alpha\beta}\,\delta(\omega+\omega') 
\,,
\label{eq:noise_xi}
\eea
where $\alpha, \beta =x, y$ denote Cartesian components.


\subsection{Active force self correlation function in $\omega$ representation}

The linearity of the dynamics for the active force, Eqs. \eqref{eqfax}-\eqref{eqfay}, allows us to solve the dynamics.
Let us introduce for convenience the dynamical matrix $\tilde{\mathbf{M}}_\bq(\omega)$ as
\begin{equation}
\tilde{\mathbf{M}}_\bq(\omega) =
\begin{bmatrix}
i \omega +\frac{1}{\tau}  & \Omega\\
-\Omega & i \omega +\frac{1}{\tau}
\end{bmatrix}
\end{equation}
and the diffusive matrix $\mathbf{D} = \mathbf{d} \mathbf{d}^T$ where $T$ means transpose matrix and $\mathbf{d}= \mathbf{I}\,m\gamma v_0/\sqrt{\tau}$.

The active force dynamics in the Fourier space \eqref{eqfax}-\eqref{eqfay} can be expressed in a matrix form as
\begin{equation}
\tilde{\mathbf{M}}_\bq(\omega) \tilde{\mathbf{f}^a}_\bq(\omega)= \sqrt{2} \mathbf{d} \cdot  \tilde{\boldsymbol{\chi}}_\bq(\omega) \,.
\label{eq:app_dyn_mat}
\end{equation} 
Dynamical correlations of the active force are defined as the elements of the matrix 
$\langle\tilde{\mathbf{f}}^a_\bq \times \left(\tilde{\mathbf{f}^a}_\bq\right)^T\rangle$, where $T$ means transpose and $\langle \cdot\rangle$ is the average over the noise realizations. 
These dynamical correlations can be obtained by multiplying the dynamics \eqref{eq:app_dyn_mat} by $\tilde{\mathbf{M}}^{-1}_\bq(\omega)$ on the left and by $\left(\tilde{\mathbf{f}_\bq^a}\right)^T$ on the right and then taking the average over the noise realizations.
Indeed, in this way, we get
\begin{equation}
\langle\tilde{\mathbf{f}_\bq^a} \times \left(\tilde{\mathbf{f}_\bq^a}\right)^T\rangle=\Bigl\langle \tilde{\mathbf{M}}^{-1}_\bq(\omega) \cdot \left[ \sqrt{2}\boldsymbol{d} \cdot \tilde{\boldsymbol{\chi}}_\bq(\omega) \times \left(\tilde{\mathbf{f}_\bq^a}\right)^T  \right] \Bigr\rangle = 2\,  \tilde{\mathbf{M}}^{-1}_\bq(\omega) \cdot \mathbf{D} \cdot \left[ \tilde{\mathbf{M}}^{-1}_{-\bq}(\omega') \right]2\pi \delta(\omega+\omega') \,.
\end{equation}
By following this strategy, we can obtain the elements of the dynamical correlations which are reported below
\begin{subequations}
  \begin{eqnarray}
&&
\langle  \tilde f_\bq^{a,x}(\omega) \tilde f_{-\bq}^{a,x}(\omega') \rangle=\frac{\frac{2 }{\tau}( m\gamma  v_0 )^2}{ [-\omega^2 +\frac{1}{\tau^2} +\Omega^2]^2 +4\frac{\omega^2}{\tau^2}}\Bigl[\Bigl( \omega^2 +\frac{1}{\tau^2} \Bigr) +\Omega^2    \Bigr] \,2\pi\delta(\omega+\omega')
\\&&
\langle  \tilde f_\bq^{a,x}(\omega) \tilde f_{-\bq}^{a,y}(\omega') \rangle=\frac{\frac{2 }{\tau}(  m\gamma v_0 )^2}{ [-\omega^2 +\frac{1}{\tau^2} +\Omega^2]^2 +4\frac{\omega^2}{\tau^2}}\Bigl[ 2i \omega\Omega   \Bigr]  \,2\pi\delta(\omega+\omega') \,.
 \end{eqnarray}
\label{eq:app_dyn_corr_active}
\end{subequations}
where the remaining elements satisfy the following relations 
\begin{flalign}
&\langle  \tilde f_\bq^{a,x}(\omega) \tilde f_{-\bq}^{a,x}(-\omega) \rangle=\langle  \tilde f_\bq^{a,y}(\omega) \tilde f_{-\bq}^{a,y}(-\omega) \rangle\\
&\langle  \tilde f_\bq^{a,x}(\omega) \tilde f_{-\bq}^{a,y}(-\omega) \rangle=- \langle  \tilde f_\bq^{a,y}(\omega) \tilde f_{-\bq}^{a,x}(-\omega) \rangle \,.
\end{flalign}


\subsection{Displacement-displacement correlation function in $\omega$ representation}

In a similar way, we can solve the linear dynamics for the particle displacement, i.e. Eqs.~\eqref{uqomegaequation} and~\eqref{underdampG} (or equivalently Eq.~\eqref{eq:dyn_disp_app}).
By multiplying the dynamics \eqref{eq:dyn_disp_app} by $\tilde u^x_{-\bq}(-\omega)$ or $\tilde u^y_{-\bq}(-\omega)$ and taking the average over the noise realization, we can express the dynamical correlations of the particle displacement
%
\begin{eqnarray}&&
\langle  \tilde u^x_{\bq}(\omega)\tilde u^x_{-\bq}(-\omega) \rangle=\tilde G_\bq ( \omega)\tilde G_{-\bq} ( -\omega)
\Bigl( \frac{1}{m^2} \langle  \tilde f_\bq^{a,x}(\omega) \tilde f_{-\bq}^{a,x}(-\omega) \rangle+2 \frac{\gamma T }{m }
\langle \xi^x_\bq(\omega) \xi^x_{-\bq}(-\omega) \rangle\Bigr)
\\&&
\langle  \tilde u_\bq^x(\omega) \tilde u^y_{-\bq}(-\omega) \rangle=\tilde G_\bq ( \omega) \tilde G_{-\bq} ( -\omega)
 \frac{1}{m^2} \langle  \tilde f_\bq^{a,x}(\omega) \tilde f_{-\bq}^{a,y}(-\omega) \rangle     
\label{offdiagonalomega}
\end{eqnarray}
Finally, we write the ($\bq,\omega$) representation of the displacement-displacement correlation function
\begin{eqnarray}&&
\label{eq:app_dyn_corr_explicit}
\langle  \tilde u^x_{\bq}(\omega)\tilde u^x_{-\bq}(\omega') \rangle=\frac{1}{(\omega^2-\omega_q^2 )^2 + \omega^2\gamma^2   }
\Bigl( 2 \frac{ \gamma T}{m }+\frac{2 v_0^2 \gamma^2}{\tau}\frac{\Bigl( \omega^2 +\frac{1}{\tau^2} \Bigr) +\Omega^2  }{ (-\omega^2 +\frac{1}{\tau^2} +\Omega^2)^2 +4\frac{\omega^2}{\tau^2}}
\Bigr) \,2\pi\delta(\omega+\omega')
\\&&
\langle  \tilde u^x_{\bq}(\omega)\tilde u^y_{-\bq}(\omega') \rangle= \frac{2  v_0^2}{\tau}\frac{\gamma^2}{( \omega^2- \omega_q^2 )^2 + \omega^2\gamma^2   }
\frac{  2i \omega\Omega       }{ (-\omega^2 +\frac{1}{\tau^2} +\Omega^2)^2 +4\frac{\omega^2}{\tau^2}}\,2\pi\delta(\omega+\omega')  \,.
\label{C24}
\end{eqnarray}
Again, the remaining matrix elements satisfy the relations
\begin{flalign}
&\langle  \tilde u^x_{\bq}(\omega)\tilde u^x_{-\bq}(-\omega) \rangle =\langle  \tilde u^y_{\bq}(\omega)\tilde u^y_{-\bq}(-\omega) \rangle\\
&\langle  \tilde u^x_{\bq}(\omega)\tilde u^y_{-\bq}(-\omega) \rangle =-\langle  \tilde u^y_{\bq}(\omega)\tilde u^x_{-\bq}(-\omega) \rangle \,.
\end{flalign}
Equations~\eqref{eq:app_dyn_corr_explicit} and~\eqref{C24} correspond to Eq.~\eqref{uxuxomega} and Eq.~\eqref{uxuyomega}, respectively. 

\subsection{Spectral representation of the active torque}

We first demonstrate that the spectral density of the angular momentum is proportional to the spectral density of the torque generated by the active force.
By multiplying the displacement dynamics \eqref{eq:dyn_disp_app} by $\tilde \bu_{-\bq}(-\omega)$ via the cross product and averaging over the noise realizations, we have:
\bea&&
\Bigl(-\omega^2+i \omega\gamma  
+  \omega_{\bq}^2 \Bigr) \langle\tilde \bu_\bq(\omega)\times \tilde \bu_{-\bq}(-\omega)\rangle= \frac{\langle\tilde {\bf f}^a_\bq(\omega)\times \tilde \bu_{-\bq}(-\omega)\rangle}{m} 
\eea
where we have used that the correlation between noise and displacement vanish.
By summing to this equation its complex conjugate, we have
\bea&&
i \omega\gamma   \langle\tilde \bu_\bq(\omega)\times \tilde \bu_{-\bq}(-\omega)\rangle + c.c=   \frac{\langle\tilde {\bf f}^a_\bq(\omega)\times \tilde \bu_{-\bq}(-\omega)\rangle}{m} + c.c. \,,
\eea
and, finally, by substituting $-i \omega \tilde \bu_{-\bq}(-\omega)\to \tilde \vv_{-\bq}(-\omega)$, we obtain the desired proportionality between angular momentum and torque:
\bea
\gamma \langle \tilde \bu_\bq(\omega) \times \tilde \vv_{-\bq}(-\omega)+c.c  \rangle= \langle \tilde  \bu_\bq(\omega) \times\tilde {\bf f}^a_{-\bq}(-\omega)+c.c  \rangle \,.
\eea
The torque generated by the active force can be calculated by multiplying the dynamics \eqref{eq:dyn_disp_app} by $\tilde{\boldsymbol{ f}}^a_{-\bq}(\omega')$ and taking the average over the noise realizations.
Within this protocol, we obtain
\bea
Re\Bigl[\langle \tilde \bu_\bq(\omega)\times \tilde{\boldsymbol{ f}}^a_{-\bq}(\omega')\rangle\Bigr]=
\frac{1}{2 m}\Bigl(\tilde G_\bq ( \omega)
\Bigl\langle  \tilde f_\bq^{a,x}(\omega) \tilde f_{-\bq}^{a,y}(-\omega)  -
  \tilde f_\bq^{a,y}(\omega) \tilde f_{-\bq}^{a,x}(-\omega) \Bigr\rangle+c.c.\Bigr)\,2\pi\delta(\omega+\omega') \, .
\eea
where we have used that the correlations between translational noise and active forces vanish.
The torque and, thus, the angular momentum can be obtained by using the expressions for the dynamical correlations of the active force \eqref{eq:app_dyn_corr_active}.
In this way, we get
\begin{eqnarray}&&
Re[\langle \tilde \bu_\bq\omega)\times  \tilde{\boldsymbol{ f}}^a_{-\bq}(\omega')\rangle]=
\frac{8 m }{\tau}(  \gamma v_0 )^2
 \frac{   \Omega   \omega^2}
{( \omega^2- \omega_q^2 )^2 + \omega^2\gamma^2   }  
\frac{1}{ (-\omega^2 +\frac{1}{\tau^2} +\Omega^2)^2 +4\frac{\omega^2}{\tau^2}} \,2\pi\delta(\omega+\omega')
\label{C28}
\end{eqnarray}
which leads to the spectral density of the angular momentum~\eqref{eq:spec_dens_M} by diving Eq.~\eqref{C28} by $\gamma$.



\vskip20pt
\section{Time correlations}
\label{timecorrelationsmethod}

In the present Appendix, we derive the time correlations in the overdamped limit by solving in the time domain the dynamical equations of the model, whose dynamics reads:
\begin{eqnarray}
\frac{ d\hat \bu_\bq(t)}{dt}
 = -\frac{\omega_\bq^2}{\gamma} \hat \bu_\bq(t) +\frac{\hat{\bf f}^a _\bq(t)}{m\gamma}  + \sqrt{\frac{ 2  T}{m \gamma}} \hat \cchi_\bq(t)  \,.
 \label{overdampeddynamicequation}
\end{eqnarray}
We first consider the following integral representation of the solutions of Eq.~\eqref{eq:activeforce}, obtained for zero initial value of the active force:
  \begin{equation}
 \left(\begin{array}{ccccccc}
\hat f_\bq^{a,x}(t)\\ \hat f_\bq^{a,y}(t)\
\end{array}\right) =  \sqrt{\frac{1 } { 2\tau}} m\gamma v_0 \int_0^t dt'
\left(\begin{array}{ccccccc}
 [e^{\lambda (t-t')}+ e^{\lambda^* (t-t')}] \hat \chi_\bq^x(t')   +i [( e^{\lambda (t-t')} - e^{\lambda^* (t-t')})
  ] \hat \chi_\bq^y(t')
\\ -i[ e^{\lambda (t-t')}-e^{\lambda^* (t-t')}]  \hat \chi_\bq^x(t') +  [e^{\lambda (t-t')}+   e^{\lambda^* (t-t')}] \hat \chi_\bq^y(t')
\end{array}\right)
\label{activeforceintegralformula}
\end{equation}
where $\lambda=-\frac{1}{\tau}+i\Omega$ and 
 $\lambda^*$ is its complex conjugate.
The solution of the overdamped equation~\eqref{overdampeddynamicequation} is given by
\begin{equation}
\hat \bu_\bq(t)=e^{-\frac{\omega_\bq^2}{\gamma} t}\, \hat \bu_\bq(0)
+\sqrt{\frac{ 2 T}{m \gamma }}\,e^{-\frac{\omega_\bq^2}{\gamma} t}\, \int_0^t \,  dt_1 \,
e^{\frac{\omega_\bq^2}{\gamma} t_1}\,
\, \hat{\cchi}_\bq^{a}(t_1)
+ e^{-\frac{\omega_\bq^2}{\gamma} t}\, \int_0^t \,  dt_1 \,
e^{\frac{\omega_\bq^2}{\gamma} t_1}\,
\, \frac{\hat{\bf f}_\bq^{a}(t_1)}{m \gamma } \,
\label{displacementintegralformula}
\end{equation}
where the first and second term represent the homogeneous part of the solution and the contribution
due to the thermal noise and the last term the effect of the activity.

\subsection{Displacement-displacement time correlation}

Putting together Eqs.~\eqref{activeforceintegralformula} and~\eqref{displacementintegralformula}
we obtain the following expression for the  part of the displacement due to the chiral active force:
\begin{equation}
  \left(\begin{array}{ccccccc}
\hat u_\bq^x(t)\\ \hat u_\bq^y(t)
\end{array}\right)_{active} =    
\frac{ v_0 }   { \sqrt{ 2\tau}} e^{-\frac{\omega_\bq^2}{\gamma} t}
 \int_0^t 
  e^{\frac{\omega_\bq^2}{\gamma} t_1}\,
   dt_1  \int_0^{t_1} dt_2
\left(\begin{array}{ccccccc}
 [e^{\lambda (t_1-t_2)}+ e^{\lambda^* (t_1-t_2)}]\, \hat{\chi}^x_\bq(t_2)   +i [( e^{\lambda (t_1-t_2)} - e^{\lambda^* (t_1-t_2)})
  ]\,\hat{\chi}^y_\bq(t_2)
\\ -i[ e^{\lambda (t_1-t_2)}-e^{\lambda^* (t_1-t_2)}] \, \hat{\chi}^x_\bq(t_2) +  [  e^{\lambda (t_1-t_2)}   +e^{\lambda^* (t_1-t_2)} ] \,\hat{\chi}^y_\bq(t_2)
\end{array}\right)\,  \nonumber
\end{equation}
that can be rewritten as
\begin{equation}
 \left(\begin{array}{ccccccc}
\hat u_\bq^x(t)\\ \hat u_\bq^y(t)
\end{array}\right)_{active}  =    \frac{  v_0 } { \sqrt{ 2\tau}}
\left(\begin{array}{ccccccc}
 [ K^x(t;\lambda)+  K^x(t;\lambda^*)]  +i [K^y(t;\lambda) - K^y(t;\lambda^*)
  ]
\\ -i[ K^x(t;\lambda) -K^x(t;\lambda^*)]  +  [K^y(t;\gamma,\lambda)+   K^y(t;\gamma,\lambda^*)] 
\end{array}\right)\,  \label{integralformula}
\end{equation}
where we used the definition
\begin{equation}
K^\alpha(t,\lambda)\equiv
e^{-\frac{\omega_\bq^2}{\gamma} t}\ \int_0^t e^{\frac{\omega_\bq^2}{\gamma} t_1}\ dt_1  \int_0^{t_1} dt_2 e^{\lambda (t_1-t_2)}\hat{\chi}^\alpha_\bq(t_2) \, .
\end{equation}
After  an integration by parts, we may rewrite it as a simple integral
\be
K^\alpha(t;\lambda)\equiv
\frac{1}{\gamma+\lambda }\int_0^t dt_1 [ e^{\lambda (t- t_1)}  - e^{-\frac{\omega_\bq^2}{\gamma} (t-t_1)} ] \hat{\chi}^\alpha_\bq(t_1)\,.
\ee
For subsequent applications, we define the function $H^\alpha(t;\lambda)$:
\begin{equation}
H^\alpha(t;\lambda)\equiv \int_0^{t} dt_1 \,e^{\lambda (t-t_1)} \hat \chi_\bq^\alpha(t_1)  \,.
\label{Hdefinition}
\end{equation}
With the help of formula~\eqref{integralformula} we can write the correlators as
\begin{equation}
\langle   \hat u_\bq^x(t) \hat u_{-\bq}^x (t')\rangle_{active} =  \frac{v^2_0} { 2 \tau} 
\Bigl\langle  K^x(t;\lambda)K^x(t';\lambda^*)+ K^x(t;\lambda^*)K^x(t';\lambda)
+  K^y(t;\lambda)K^y(t';\lambda^*)+ K^y(t;\lambda^*)K^y(t';\lambda)
 \Bigr\rangle
 \label{eqC8}
\end{equation}
where the angular brackets represent the double average over the thermal noise and the realisations of 
active noise. 
using the following properties of the averages $\langle  K^x(t;\lambda)K^x(t';\lambda^*)\rangle=\langle  K^y(t;\lambda)K^y(t';\lambda^*)\rangle$
and $\langle  K^x(t;\lambda)K^x(t';\lambda)\rangle=\langle  K^y(t;\lambda)K^y(t';\lambda)\rangle$,
we find
\bea&&
\langle K^x(t;\lambda)K^x(t';\lambda^*)\rangle= \frac{1}{[\frac{\omega_\bq^2}{\gamma}-\frac{1}{\tau}]^2+\Omega^2 }\nonumber\\&&
\Bigl(  -\frac{1}{\lambda +\lambda^*}  [e^{ \lambda_1 (t-t')} -e^{\lambda t+\lambda^*t'}] + \frac{1}{2\frac{\omega_\bq^2}{\gamma}} [ e^{-\frac{\omega_\bq^2}{\gamma}(t- t')}-e^{-\frac{\omega_\bq^2}{\gamma} (t+t')}    ] - \frac{1}{\frac{\omega_\bq^2}{\gamma} -\lambda}  [e^{\lambda (t-t')} -e^{\lambda t-\frac{\omega_\bq^2}{\gamma} t'}]\nonumber\\&&
- \frac{1}{\frac{\omega_\bq^2}{\gamma} -\lambda^*}  [e^{-\frac{\omega_\bq^2}{\gamma}(t- t')} -e^{\lambda^* t'-\frac{\omega_\bq^2}{\gamma} t}] 
\Bigr)\,.
\label{eqC9}
\eea
Performing the calculations after simple algebra we find the result of Eq.~\eqref{diagonaluu}.
Similarly, we obtain the off diagonal correlation function:
\beq
\langle   \hat u_\bq^x(t) \hat u_{-\bq}^y (t')\rangle_{active}  = i \frac{   v^2_0} {  2\tau} 
\Bigl\langle  [ K^x(t;\lambda)K^x(t';\lambda^*)- K^x(t;\lambda^*)K^x(t';\lambda)] +
[ K^y(t;\lambda)K^y(t';\lambda^*)- K^y(t;\lambda^*)K^y(t';\lambda)] 
 \Bigr\rangle
 \label{eqC10}
\eeq
Explicitly:
\bea&&
 \Bigl\langle  K^x(t;\lambda)K^x(t';\lambda^*)- K^x(t;\lambda^*)K^x(t';\lambda)
 \Bigr  \rangle= 
 \frac{1}{[\frac{\omega_\bq^2}{\gamma}-\frac{1}{\tau}]^2+\Omega^2 }
 \Bigl(   -\frac{1}{\lambda +\lambda^*}   [e^{ \lambda (t-t')}-e^{ \lambda^* (t-t')} ]\nonumber\\&&
 - \frac{1}{\frac{\omega_\bq^2}{\gamma} -\lambda}  e^{\lambda (t-t')} 
  + \frac{1}{\frac{\omega_\bq^2}{\gamma} -\lambda^*}  e^{\lambda^* (t-t')}      %
  - \frac{1}{\frac{\omega_\bq^2}{\gamma} -\lambda^*}  e^{-\frac{\omega_\bq^2}{\gamma}(t- t')} 
   + \frac{1}{\frac{\omega_\bq^2}{\gamma} -\lambda}  e^{-\frac{\omega_\bq^2}{\gamma}(t- t')}
\Bigr)
\label{u1u2time}
 \eea
leading to  formula \eqref{offdiagonaluu}.
To obtain the  cross correlation, $\langle   \hat u_\bq^y(t) \hat u_{-\bq}^x (t')\rangle $, we use the property
\beq
\langle   \hat u_\bq^y(t) \hat u_{-\bq}^x (t')\rangle  
 = -\langle   \hat u_\bq^x(t) \hat u_{-\bq}^y (t')\rangle  
\eeq
from which it follows that the equal-time cross correlations vanish:
\beq
\langle   \hat u_\bq^y(t) \hat u_{-\bq}^x (t)\rangle  =
  -\langle   \hat u_\bq^x(t) \hat u_{-\bq}^y (t)\rangle  =0 \, .
\eeq

 \subsection{ Velocity-Velocity time-correlation}

 The velocity correlations can be computed by differentiating  the displacement correlation function with respect to its two arguments, $t$ and $t'$:
 \bea&&
 \langle   \hat v_\bq^x(t) \hat v_{-\bq}^x(t')\rangle_{active}  =  
\frac{d}{dt}  \frac{d}{dt'} \langle   \hat u_\bq^x(t) \hat u_{-\bq}^x (t')\rangle_{active} 
\label{vqvqcorrelationequaltime}
 \eea
and a similar method is employed to determine  the off diagonal velocity correlation. The relative results
are given in the main text Eq.~\eqref{vxvxtimecorrel} and~\eqref{eq:approx_vxvy_time}.
Evaluating Eq.\eqref{vqvqcorrelationequaltime} at $t=t'$, one obtains:
\bea&&
  \langle \hat v_\bq^\alpha(t) \hat v_{-\bq}^\alpha(t) \rangle_{active}= v^2_0
\frac{ 1+\frac{\tau}{\gamma} \omega_q^2+\Omega^2\tau^2}{(1+ \frac{\tau}{\gamma}\omega_q^2)^2+\Omega^2\tau^2} \,,
   \label{selfvelocity0bb}
\eea
which corresponds to Eq.~\eqref{equaltimevv}.

\vskip20pt
\section{Torque and angular momentum}
\label{Torquemomentum}

Let us consider 
 $\mathbf{T}$, the average total torque exerted by the active forces on the particles. Its expression is:
\begin{equation}
{\bf T}=   \sum_\bn
\langle \bu_\bn(t) \times {\bf f}^a_\bn(t)\rangle \,.
\label{Torquefa}
\end{equation}
On the other hand the average total angular momentum is
 \begin{equation}
 {\bf M}=m\sum_{\bn} \langle\bu_{\bn}(t)\times \vv_{\bn}(t) \rangle \,.
 \label{angular_momentum}
 \end{equation}


To evaluate the average angular momentum \eqref{angular_momentum}, we consider the time derivative of the displacement correlation:
\beq
\frac{d}{dt}\langle   \hat u_\bq^x(t) \hat u_{-\bq}^y (t')\rangle  = i \frac{   v^2_0} {  2\tau} 
\frac{d}{dt}\Bigl\langle  [ K^x(t;\lambda)K^x(t';\lambda^*)- K^x(t;\lambda^*)K^x(t';\lambda)] +
[ K^y(t;\lambda)K^y(t';\lambda^*)- K^y(t;\lambda^*)K^y(t';\lambda)] \,.
 \Bigr\rangle_{t=t'}
\eeq
If one  considers $t>t'$ but both $t$ and $t'$ are $\gg\tau$ and $\gg1/\gamma$ , one obtains the result:
 \bea&&
\frac{d}{dt}\Bigl\langle  K^x(t;\lambda)K^x(t';\lambda^*)- K^x(t;\lambda^*)K^x(t';\lambda)
 \Bigr\rangle= \frac{1} {(\frac{\omega_\bq^2}{\gamma}-\frac{1}{\tau})^2+\Omega^2 }\nonumber\\&&
\Bigl(   -\frac{1}{\lambda +\lambda^*}   \lambda e^{ \lambda (t-t')}-\lambda^* e^{ \lambda^* (t-t')} 
 - \frac{1}{\frac{\omega_\bq^2}{\gamma} -\lambda}  \lambda e^{\lambda (t-t')} 
  + \frac{1}{\frac{\omega_\bq^2}{\gamma} -\lambda^*}  \lambda^*e^{\lambda^* (t-t')} 
  \nonumber\\&&
  +\frac{1}{\frac{\omega_\bq^2}{\gamma} -\lambda^*}  \frac{\omega_\bq^2}{\gamma} e^{-\frac{\omega_\bq^2}{\gamma}(t- t')} 
   - \frac{1}{\frac{\omega_\bq^2}{\gamma} -\lambda}  \frac{\omega_\bq^2}{\gamma} e^{-\frac{\omega_\bq^2}{\gamma}(t- t')}
\Bigr) \,.
 \eea
 After some algebra, we obtain:
\beq
m\langle   \hat \bu_\bq(t) \times \hat \vv_{-\bq} (t)\rangle=
2 m  v^2_0\tau \Bigl( \frac{\Omega \tau}{(1+\frac{\tau\omega_\bq^2}{\gamma})^2+\Omega^2\tau^2}\Bigr) \,.
\label{angularmomentum2}
\eeq

In order to evaluate the average torque \eqref{Torquefa}
 exerted on the $\bq$ mode we consider the following cross correlation:
\beq
\langle   \hat u_\bq^x(t) \hat f^{a,y}_{-\bq} (t')\rangle  = i \frac{ m\gamma  v^2_0} {  2\tau}
\Bigl\langle  [ K^x(t;\lambda)H^x(t';\lambda^*)- K^x(t;\lambda^*)H^x(t';\lambda)] +
[ K^y(t;\lambda)H^y(t';\lambda^*)- K^y(t;\lambda^*)H^y(t';\lambda)] 
 \Bigr\rangle \, .
\eeq
Performing the integrals and averaging over realizations we find
\bea
\Bigl\langle  K^x(t;\lambda)H^x(t;\lambda^*)- K^x(t;\lambda^*)H^x(t;\lambda) \Bigr\rangle=
-\frac{1}{\frac{\omega_\bq^2}{\gamma}+\lambda } \frac{1}{\frac{\omega_\bq^2}{\gamma}+\lambda^* }  
\Bigl(\frac{\lambda^*-\lambda}{\lambda+\lambda^* }  
+\frac{\frac{\omega_\bq^2}{\gamma}+\lambda^*}{\frac{\omega_\bq^2}{\gamma}-\lambda^* }  -\frac{\frac{\omega_\bq^2}{\gamma}+\lambda}{\frac{\omega_\bq^2}{\gamma}-\lambda }  \Bigr) \,.
\eea
A brief calculation yields the following result:
\bea
\langle   \hat u_\bq^x(t) \hat f^{a,y}_{-\bq} (t) - \hat u_\bq^y(t) \hat f^{a,x}_{-\bq} (t)\rangle  
=2 m   \gamma   v^2_0 \tau
\Bigl(\frac{\Omega \tau}{  (1+\frac{\tau \omega_\bq^2}{\gamma})^2+\Omega^2  \tau^2    }  \Bigr) \,.
\label{formulav1eta2}
\eea
By comparison with Eq.~\eqref{angularmomentum2} one obtains that
 the torque exerted by the frictional forces
  represented by formula \eqref{formulav1eta2}  is equal to the angular momentum multiplied by $\gamma$.
 In other words, the torque exerted by the 
active forces is balanced by the frictional torque exerted by the medium.

\vskip20pt
\section{Derivation of the Entropy production formula}\label{app:EPR}

To use the definition \eqref{eq:EPR} and calculate the total entropy production rate $\dot{S}$, we use a path-integral approach to estimate the probability of forward and backward trajectories.
As usual, the particles' trajectories are fully determined by the noise path.
For a chiral active systems, the noise probability has a Gaussian form for each noise governing the dynamics at each time, $\boldsymbol{\xi}(t)$ and $\boldsymbol{\chi}(t)$.
By using curly bracket to denote a noise trajectory from an initial time $t_0$ to a final time $t_f$ for every particle of the system, i.e.\ $(\{\boldsymbol{\xi}\}, \{\boldsymbol{\chi}\})$, the path probability of the noise reads
\begin{equation}
\text{Prob}(\{\boldsymbol{\xi}\}, \{\boldsymbol{\chi}\}) \propto \exp{\left( -\frac{1}{2} \sum_{\bn}\int_{t_0}^{t_f} dt\, \boldsymbol{\xi}_\bn^2(t)  \right)} \exp{\left( -\frac{1}{2}  \sum_{\bn}\int_{t_0}^{t_f} dt\, \boldsymbol{\chi}^2_\bn(t)  \right)}\,.
\end{equation}
By expressing each noise as a function of the dynamical variables $\mathbf{x}_\bn$, $\mathbf{v}_\bn$, and $\mathbf{f}^a_\bn$ via by using the equation of motion
\begin{subequations}
\begin{flalign}
&\boldsymbol{\chi}_\bn=\frac{\sqrt \tau}{\sqrt{ 2} m\gamma v_0} \Bigl( \dot{\mathbf{f}}^{a}_\bn+\frac{1}{\tau} f_\bn^{a,x}-\boldsymbol{\Omega}\times \mathbf{f}_\bn^{a}  \Bigr)\\
& \xxi_\bn  = \sqrt{\frac{ m}{2 \gamma T}}  \left(\ddot \bu_\bn+
\gamma \dot \bu_\bn - \frac{{\bf f}^a _\bn}{m}  - \frac{\mathbf{F}_\bn}{m}   \right) \,,
\end{flalign}
\label{eq:app_changeofvar}
\end{subequations}
one can easily switch from the probability of the noise $\text{Prob}(\{\boldsymbol{\xi}\}$ to the probability of the trajectory $\mathcal{P}_f((\{\mathbf{v}\}, \{\mathbf{v}\}, \{\mathbf{f}^a\})$ by applying a change of variables.
Since the Jacobian of this transformation does not affect the entropy production \cite{caprini2019entropy}, we can explicitly write
\begin{flalign}
\mathcal{P}_f((\{\mathbf{u}\}, \{\mathbf{v}\}, \{\mathbf{f}^a\})\propto &\exp{\left( -\frac{ m}{4 \gamma T}  \sum_{\bn}\int_{t_0}^{t_f} dt\,  \left(\ddot \bu_\bn+
\gamma \dot \bu_\bn - \frac{{\bf f}^a _\bn}{m}  - \frac{\mathbf{F}_\bn}{m}   \right)^2  \right)} \times \nonumber\\
&\times\exp{\left( -\frac{\tau}{4 m^2\gamma^2 v_0^2}  \sum_{\bn}\int_{t_0}^{t_f} dt\,  \Bigl( \dot{\mathbf{f}}^{a}_\bn+\frac{1}{\tau} f_\bn^{a,x}-\boldsymbol{\Omega}\times \mathbf{f}_\bn^{a}  \Bigr)^2  \right)}\,.
\label{eq:app_prof_f}
\end{flalign}
The probability of the backward trajectory $\mathcal{P}_f$ can be obtained by applying the following transformations
\begin{subequations}
\begin{flalign}
&t \to -t\\
&\mathbf{x}_n\to\mathbf{x}_n\\
&\mathbf{v}_n\to -\mathbf{v}_n\\
&\mathbf{f}^a_n\to\mathbf{f}^a_n \,,
\end{flalign}
\end{subequations}
to the particle dynamics, and consequently to the relations in \eqref{eq:app_changeofvar}, which are involved in the path probability.
The assumption of even position and odd velocity under time-reversal transformation appears natural. In contrast, assuming an even active force is less straightforward and has recently sparked debate within the active matter community \cite{dabelow2019irreversibility, mandal2017entropy, caprini2019entropy, caprini2018comment}.
When calculating the log ratio of the path probabilities, each noise source contributes an additional term. Therefore, we will separately calculate the contributions to the entropy production rate arising from the thermal noise $\boldsymbol{\xi}_\bn$, governing the velocity dynamics, and the noise $\boldsymbol{\chi}_\bn$, controlling the active force dynamics.
In other words, the total entropy production rate can be decomposed into the sum of two terms
\begin{equation}
\dot{S}= \dot{S}^{v} + \dot{S}^f
\end{equation}
where the subscripts highlight the origin of the entropy production rate: $v$ for velocity dynamics and $f$ for active force dynamics.


 \subsection{Contribution of the velocity dynamics to entropy production}
 \label{EPRactiveparticle}

The Contribution of the velocity dynamics to the entropy production rate can be calculated by restricting to the first exponential in the expression \eqref{eq:app_prof_f}. We remark that this calculation is formally equivalent to the one for non-chiral particles and, indeed, it leads to the same result.
Indeed, by considering the log ratio of the forward and backward trajectories in this exponential contribution, we obtain
\begin{equation}
\dot{S}^v =\frac{1}{T}\sum_\bn \langle \mathbf{v}_\bn(t)\cdot\mathbf{f}^a_\bn(t)\rangle
\label{eq:app_Sv}
\end{equation}
where $b.t.$ denotes boundary terms that vanish in the infinite time limit, i.e.\ the potential and kinetic energy.
This term coincides with the first term in Eq. \eqref{eq:Sdot_sum}.

Calculating $\dot{S}^v$ requires to estimate the equal time correlation appearing in Eq.~\eqref{eq:app_Sv}.
Employing Eqs. \eqref{integralformula}, \eqref{activeforceintegralformula}, and \eqref{Hdefinition},
the product the velocity of the particles and the active force can be represented as:
\bea&&
\langle   \hat v_\bq^x(t) \hat f^{a,x}_{-\bq} (t')\rangle = m\gamma \frac{v^2_0} { 2 \tau} \Bigl(
\frac{d}{dt}\langle K^x(t;\lambda)H^x(t';\lambda^*)\rangle_{t=t'}+\frac{d}{dt}\langle K^x(t;\lambda^*)H^x(t';\lambda)
\rangle_{t=t'}+\nonumber\\&&
\frac{d}{dt}\langle K^y(t;\lambda)H^y(t';\lambda^*)\rangle_{t=t'}+\frac{d}{dt}\langle K^y(t;\lambda^*)H^y(t';\lambda)
\rangle_{t=t'}
 \Bigr) \,.
 \eea
When $t\to \infty$ and $t'\to \infty$ but $(t-t')\geq 0$ remains finite so that for $t=t'$ we obtain:
\bea&&
\frac{d}{dt}\langle K^x(t;\lambda)H^x(t';\lambda^*)\rangle_{t=t'}+\frac{d}{dt}\langle K^x(t;\lambda^*)H^x(t';\lambda)\rangle_{t=t'}\nonumber\\&&
=
\frac{1}{\frac{\omega_\bq^2}{\gamma}+\lambda }  
\Bigl(  -\frac{\lambda}{\lambda +\lambda^*} 
 +\frac{\frac{\omega_\bq^2}{\gamma}}{\frac{\omega_\bq^2}{\gamma} -\lambda^*}  
\Bigr)
+\frac{1}{\frac{\omega_\bq^2}{\gamma}+\lambda^* }  
\Bigl(  -\frac{\lambda^*}{\lambda +\lambda^*} 
 +\frac{\frac{\omega_\bq^2}{\gamma}}{\frac{\omega_\bq^2}{\gamma} -\lambda}\Bigr)  
= \frac{\frac{\omega_\bq^2}{\gamma}+\frac{1}{\tau}+\Omega^2\tau}{(\frac{\omega_\bq^2}{\gamma}+\frac{1}{\tau})^2+\Omega^2}
\eea
and use the property of the averages:
$\langle  K^x(t;\lambda)H^x(t';\lambda^*)\rangle=\langle  K^y(t;\lambda)H^y(t';\lambda^*)\rangle$.
Finally, we find
\bea&&
\langle   \hat v_\bq^x(t) \hat f^{a,x}_{-\bq} (t)\rangle = m\gamma v^2_0 
 \frac{1+\frac{\tau \omega_\bq^2}{\gamma}+\Omega^2\tau^2}{(1+\frac{\tau \omega_\bq^2}{\gamma})^2+\Omega^2\tau^2} \,,
\eea
which reduces to $\langle   \hat v_\bq^x(t) \hat f^{a,x}_{-\bq} (t)\rangle= m\gamma v^2_0  
(1+\frac{\tau \omega_\bq^2}{\gamma})^{-1}$ in the achiral case, $\Omega=0$.
The entropy rate production of the mode $\bq$ is given by
\bea&&
\frac{1}{T}\Bigl\langle   \hat v_\bq^x(t) \hat f^{a,x}_{-\bq} (t)+   \hat v_\bq^y(t) \hat f^{a,y}_{-\bq} (t)\Bigr\rangle = 2   \frac{m\gamma v^2_0  }{T}
 \frac{1+\frac{\tau \omega_\bq^2}{\gamma}+\Omega^2\tau^2}{(1+\frac{\tau \omega_\bq^2}{\gamma})^2+\Omega^2\tau^2}\,.
\eea
From here, we can calculate the velocity-active force correlation in real space, and thus entropy production, by integrating over the wave vector $\bq$, as done in Sec.~\ref{Staticcorrelations}.
This leads to the following expression for the velocity contribution to the entropy production rate of the system $\dot{S}^v$:
\begin{eqnarray}&&
\dot S^v\approx  \frac{v_c}{4\pi} \frac{m\gamma^2 v^2_0  }{T} \frac{1}{  \tau c^2}
\Bigl\{ 
 \ln\Bigl(\frac{(1+\frac{\tau}{\gamma} c^2 q_D^2)^2+\Omega^2\tau^2 }{  1+\Omega^2\tau^2   }\Bigr)+2 |\Omega| \tau  \arctan\Bigl( \frac{  \frac{ \frac{\tau}{\gamma} c^2 q_D^2  }{|\Omega| \tau }}
{1+ \frac{ 1+ \frac{\tau}{\gamma} c^2 q_D^2  }{\Omega^2 \tau^2 }}
  \Bigr)  
\Bigr\}
\label{Bmagnlatticed3b}
\end{eqnarray}
which corresponds to Eq.~\eqref{magnlatticed3b}.
 
  \subsection{Contribution of the active force dynamics to entropy production}
\label{EPRactivebath}

In this section, we calculate the contribution to entropy production of the active force dynamics, i.e.\ $\dot{S}^f$.
To increase the clarity of this calculation, here we report the path probabilities related to the active force dynamics in Cartesian components
\beq
\mathcal{P}_f\propto \exp\Biggl(-\frac{\tau}{(2 m\gamma v_0)^2} \sum_\bn \int dt  \Bigl( \dot{f}^{a,x}_\bn+\frac{1}{\tau} f_\bn^{a,x}+\Omega  f_\bn^{a,y}  \Bigr)^2  -  \frac{\tau}{(2 m\gamma v_0)^2} \sum_\bn\int dt  \Bigl( \dot{f}^{a,y}_\bn+\frac{1}{\tau} f_\bn^{a,y}-\Omega  f_\bn^{a,x}  \Bigr)^2     \Biggr)\,.
\eeq
In agreement with previous work, we assume that the active force is even under time-reversal transformation, such that $\mathbf{f}_\bn^{a}\to \mathbf{f}_\bn^{a}$ when $t\to-t$.
The probability of the reversed path $\mathcal{P}_r$ associated to the active force dynamics reads:
\begin{equation}
\mathcal{P}_r \propto  \exp\Biggl(-\frac{\tau}{(2 m\gamma v_0)^2} \sum_\bn\int dt  \Bigl(- \dot{f}^{a,x}_\bn+\frac{1}{\tau} f_\bn^{a,x}+\Omega  f_\bn^{a,y}  \Bigr)^2  -  \frac{\tau}{(2m\gamma v_0)^2} \sum_\bn \int dt  \Bigl(- \dot{f}^{a,y}_\bn+\frac{1}{\tau} f_\bn^{a,y}-\Omega  f_\bn^{a,x}  \Bigr)^2     \Biggr)\,.
\end{equation}
By taking the ratio $\ln{(\mathcal{P}_f/\mathcal{P}_r)}$, applying the average and dividing by the time interval, we obtain the entropy production definition.
Since noises of different particles are independent we have:
\begin{equation}
\begin{aligned}
\dot S^f &=\sum_{\bn}^N \dot s_\bn^f
\end{aligned}
\end{equation}
where $\dot{s}_\bn^f$ is the entropy production of the medium per particle, that is
\begin{equation}
\begin{aligned}
\dot s^f_\bn&=- \frac{\tau}{(m \gamma v_0)^2} \int_{t_0}^{t_f} dt 
  \dot{f}^{a,x}_\bn\left(\frac{1}{\tau} f_\bn^{a,x}+\Omega  f_\bn^{a,y}   \right)
-\frac{\tau}{(m \gamma v_0)^2}\int_{t_0}^{t_f} dt  \dot{f}^{a,y}_\bn  \left( \frac{1}{\tau} f_\bn^{a,y}-\Omega  f_\bn^{a,x}    \right)\,.
\end{aligned}
\end{equation}
In the previous expression, only cross terms $xy$ survive providing a contribution to entropy production rate per particle which reads
\begin{flalign}
\dot s^f_\bn&=- \frac{ \Omega \tau}{(m \gamma v_0)^2}\,\lim_{(t_f -t_0 )\to \infty} \,\frac{1}{(t_f - t_0 )}\int_{t_0}^{t_f} dt 
\left(  \dot{f}^{a,x}_\bn f_\bn^{a,y}   - \dot{f}^{a,y}_\bn    f_\bn^{a,x}    \right)\nonumber\\
&= - \frac{ \Omega \tau}{(m \gamma v_0)^2} \left\langle  \dot{f}^{a,x}_\bn f_\bn^{a,y}   - \dot{f}^{a,y}_\bn    f_\bn^{a,x}    \right\rangle\,,
\end{flalign}
where $t=t_f-t_0$.
Finally, by using Eqs. \eqref{exey} and \eqref{e_cross}
\bea&&
\frac{\partial }{\partial t}\langle f_\bn^{a,x}(t) f_{\bn'}^{a,y}(t')\rangle_{t=t'} 
=-  m^2\gamma^2 v_0^2 \, \delta_{\bn,\bn'} \Omega  
\eea
we obtain the final result which has the following form in the infinite time limit, i.e.\ for $(t_f - t_0 )\to \infty$:
\beq
\dot S^f=\frac{1}{N}\sum_{\bn}\dot s^f_\bn=2\Omega^2\tau \,.
\eeq
This expression gives the contribution of the active bath to the entropy production rate reported in Eq.~\eqref{EPRchiralbath}.

\end{strip}

\section*{Conflicts of interest}
There are no conflicts to declare.

\section*{Acknowledgements}
LC and UMBM acknowledge Andrea Puglisi for useful discussions.





\bibliography{bibliobozza} 
\bibliographystyle{rsc} 

\end{document}